\documentclass[conference]{IEEEtran}
\IEEEoverridecommandlockouts
\usepackage{cite}
\usepackage{amsmath,amssymb,amsfonts}
\usepackage{algorithmic}
\usepackage{graphicx}
\usepackage{textcomp}
\usepackage{xcolor}
\usepackage{graphicx}
\usepackage{dcolumn}
\usepackage{bm}
\usepackage{cuted}
\usepackage{xcolor}
\usepackage[utf8]{inputenc}
\usepackage[T1]{fontenc}
\usepackage{mathptmx}
\usepackage{amsmath}

\newcommand{\ket}[1]{\left|#1\right\rangle}
\newcommand{\bra}[1]{\left\langle#1\right|}
\newcommand{\braket}[2]{\left\langle#1\middle|#2\right\rangle}

\def\BibTeX{{\rm B\kern-.05em{\sc i\kern-.025em b}\kern-.08em
    T\kern-.1667em\lower.7ex\hbox{E}\kern-.125emX}}
\begin{document}

\title{Towards Quantum Internet and Non-Local Communication in Position-Based Qubits \\
}

\author{\IEEEauthorblockN{1\textsuperscript{st} Krzysztof Pomorski$^{a,b}$}
\IEEEauthorblockA{\textit{University College Dublin, Dublin, Ireland} \\ 
\textit{a: School of Computer Science}\\ 
\textit{b: School of Electrical and Electronic Engineering}\\
\textit{Dublin, Ireland}  \\
\textit{c: Quantum Hardware Systems} \\
Webpage: \textit{www.quantumhardwaresystems.com} \\
E-mail: \textit{kdvpomorski@gmail.com}}
\and
\IEEEauthorblockN{2\textsuperscript{nd} Robert Bogdan Staszewski}
\IEEEauthorblockA{\textit{University College Dublin} \\
\textit{School of Electrical and Electronic Engineering}\\
Dublin, Ireland }

}

\maketitle

\begin{abstract}
Non-local communication between position-based qubits is described for a system of a quantum electromagnetic resonator entangled to two semiconductor electrostatic qubits via an interaction between matter and radiation by Jaynes-Cummings tight-binding Hamiltonian. Principle of quantum communication between position-dependent qubits is explained. Further prospects of the model development are given. The obtained results bring foundation for the construction of quantum internet and quantum communication networks between position-based qubits that are implementable in semiconductor single-electron devices that can be realized in current CMOS technologies.
The case of two semiconductor
position-dependent qubits interacting with quantum electromagnetic cavity is
discussed and general form of tight-binding Hamiltonian is derived with
renormalized tight-binding coefficients. The considerations also describe the situation of mutual qubit-qubit electrostatic interaction.
\end{abstract}


\section{Technological motivation
}
\begin{figure}
    \centering
    \includegraphics[width=0.95\columnwidth]{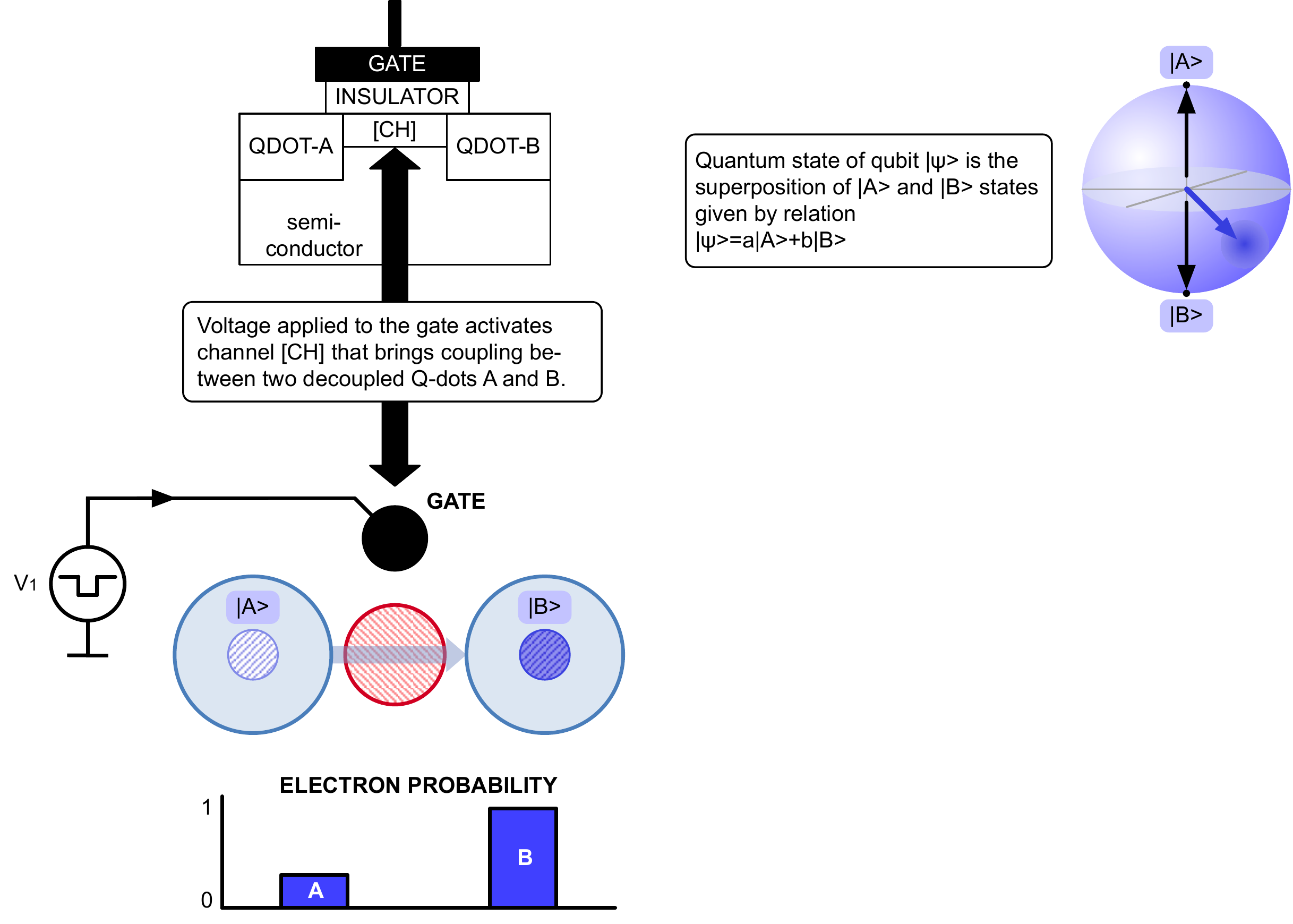} 
    \caption{Basic concept of position based qubit \cite{Pomorski_spie} and its correspondence to Bloch sphere \cite{Nbody}.}
 \label{central1}
\end{figure}
Single-electron semiconductor devices are now actively researched for their potential in realizing quantum computers (QC), and especially for implementing single-chip CMOS QCs that are fully integrated with their surrounding electronics \cite{Bashir19}. They were studied by Fujisawa \cite{Fujisawa}, Petta \cite{Petta}, Leipold \cite{Dirk}, Giounanlis \cite{Panos}, Pomorski \cite{Pomorski_spie} and many others. On the other hand, one of the most successful models in condensed matter physics is Hubbard model and its special case known as tight-binding model \cite{Spalek}. We consider a two-energy-level system of position-based (a.k.a. charge) qubit in a tight-binding approach that is a predecessor of Hubbard model as depicted in Fig.\,\ref{central1}. 

The Hamiltonian of this system
is given as 
\begin{strip}
\begin{eqnarray}
\label{simplematrix}
\hat{H}(t)_{[x=(x_1,x_2)]}=
\begin{pmatrix}
E_{p1}(t) & t_{s12}(t)=|t_{s12}|e^{+i\alpha(t)} \\
t_{s12}^{\dag}(t)=|t_{s12}|e^{-i\alpha(t)} & E_{p2}(t)
\end{pmatrix}=\nonumber \\
E_{p1}(t)\ket{x_1}\bra{x_1}+E_{p2}(t)\ket{x_2}\bra{x_2}+t_{s12}(t)\ket{x_1}\bra{x_2}+t_{s21}(t)\ket{x_2}\bra{x_1}
=(E_1(t)\ket{E_1}_t \bra{E_1}_t+E_2(t)\ket{E_2}\bra{E_2})_{[E=(E_1,E_2)]}.
\end{eqnarray}
\end{strip}
The $\hat{H}(t)$ Hamiltonian's eigenenergies $E_1(t)$ and $E_2(t)$, with $E_2(t)>E_1(t)$, are given,
with $t_{s12}(t)=t_{sr}(t)+it_{si}(t)$, as:
\begin{eqnarray*}
E_1(t)= \left(-\sqrt{\frac{(E_{p1}(t)-E_{p2}(t))^2}{4}+|t_{s12}(t)|^2}+ \frac{E_{p1}(t)+E_{p2}(t)}{2}\right), 
E_2(t)= \left(+\sqrt{\frac{(E_{p1}(t)-E_{p2}(t))^2}{4}+|t_{s12}(t)|^2}+\frac{E_{p1}(t)+E_{p2}(t)}{2}\right), 
\end{eqnarray*}
and energy eigenstates $\ket{E_1(t)}$ and $\ket{E_2(t)}$ are expressed in terms of maximum localized state on the left and right node and have the following form
\onecolumn
\begin{eqnarray}
\ket{E_1,t}=
\begin{pmatrix}
\frac{(E_{p2}(t)-E_{p1}(t))+\sqrt{\frac{(E_{p2}(t)-E_{p1}(t))^2}{2}+|t_{s12}(t)|^2}}{-i t_{sr}(t)+t_{si}(t)} \\
-1
\end{pmatrix}=
\frac{(E_{p2}(t)-E_{p1}(t))+\sqrt{\frac{(E_{p2}(t)-E_{p1}(t))^2}{2}+|t_{s12}(t)|^2}}{-i t_{sr}(t)+t_{si}(t)}\ket{x_1}-\ket{x_2},\nonumber \\
\ket{E_2,t}=
\begin{pmatrix}
\frac{-(E_{p2}(t)-E_{p1}(t))+\sqrt{\frac{(E_{p2}(t)-E_{p1}(t))^2}{2}+|t_{s12}(t)|^2}}{t_{sr}(t) - i t_{si}(t)} \\
1
\end{pmatrix}=
\frac{-(E_{p2}(t)-E_{p1}(t))+\sqrt{\frac{(E_{p2}(t)-E_{p1}(t))^2}{2}+|t_{s12}(t)|^2}}{t_{sr}(t) - i t_{si}(t)}\ket{x_1}+\ket{x_2}. \nonumber \\
\end{eqnarray}
%
The last expressions can be written in a compact form
\begin{eqnarray}
\begin{pmatrix}
\ket{E_1,t} \\
\ket{E_2,t} \\
\end{pmatrix}=
\hat{S}_{2 \times 2}
\begin{pmatrix}
\ket{x_1} \\
\ket{x_2} \\
\end{pmatrix}
=
\begin{pmatrix}
\frac{(E_{p2}(t)-E_{p1}(t))+\sqrt{\frac{(E_{p2}(t)-E_{p1}(t))^2}{2}+|t_{s12}(t)|^2}}{-i t_{sr}(t)+t_{si}(t)} & -1 \\
\frac{-(E_{p2}(t)-E_{p1}(t))+\sqrt{\frac{(E_{p2}(t)-E_{p1}(t))^2}{2}+|t_{s12}(t)|^2}}{t_{sr}(t) - i t_{si}(t)} & 1 \\
\end{pmatrix}
\begin{pmatrix}
\ket{x_1} \\
\ket{x_2} \\
\end{pmatrix}.
\end{eqnarray}
Setting $t_{si}(t)=1, t_{sr}(t)=0$ and $E_{p1}(t)=E_{p2}(t)=E_p$ we obtain
$\begin{pmatrix}
\ket{E_2}_n \\
\ket{E_1}_n \\
\end{pmatrix}=\frac{1}{\sqrt{2}}
\begin{pmatrix}
1 &  +1 \\
1 &  -1 \\
\end{pmatrix}
\begin{pmatrix}
\ket{x_2} \\
\ket{x_1} \\
\end{pmatrix}$
which brings Hadamard matrix as relating q-state in the position base and in the energy base, where $\ket{E_{1(2)}}_n=\frac{1}{\sqrt{2}}\ket{E_{1(2)}}$.
If we associate logic state 0 with occupancy of node 1 (spanned by $\ket{x_1}$) and logic state 1 with occupancy of node 2 spanned by $\ket{x_2}$, then Hadamard operation on logic state $0$ brings occupancy of $E_2$ (so it is spanned by $\ket{E_2}$) and Hadamard operation on
logic state $1$ brings the entire occupancy of energy level $E_1$ (that is spanned by $\ket{E_1}$).

It shall be underlined that in the most simple case of position-based qubit $E_{p1}=E_{p2}=E_p=\rm const_1$ and $t_{s12}=|t|=\rm const_2$ and we obtain
%
$\ket{\psi(t)}=\frac{1}{\sqrt{2}}(c_{E_1}e^{\frac{E_1}{\hbar}t}+c_{E2}e^{\frac{E_2}{\hbar}t})\ket{x_1}
+\frac{1}{\sqrt{2}}(-c_{E1}e^{\frac{E_1}{\hbar}t}+c_{E_2}e^{\frac{E_2}{\hbar}t})\ket{x_2}$.
%
It implies an oscillation of probabilities for the electron presence at node 1 (quantum logical 0) and 2 (quantum logical 1) with frequency $2|t|=E_2-E_1$, where $|c_{E_1}|^2 (|c_{E_2}|^2)$ is the probability for the quantum state to be in the ground (excited) state. 
It is possible to determine the qubit state under any evolution of two eigenergies $E_1(t)$ and $E_2(t)$ that are dependent on $E_{p1}(t), E_{p2}(t), t_{s12}(t)=t_{sr}(t)+t_{si}(t)i$. Simply, we have the state at any time instant given by \begin{eqnarray}
\ket{\psi_t}=e^{\int_{t_0}^{t}\frac{1}{\hbar i}\hat{H}(t')dt'}\ket{\psi_{t_0}}=\hat{U}(t,t_0)\ket{\psi_{t_0}}=
\begin{pmatrix}
e^{\frac{1}{\hbar i}\int_{t_0}^{t}E_1(t')dt'}, 0 \\
0 & e^{\frac{1}{\hbar i}\int_{t_0}^{t}E_2(t')dt'} \\
\end{pmatrix}\ket{\psi_{t_0}},
\end{eqnarray}
We notice that in case of qubit the evolution operator is given as
\begin{eqnarray}
\hat{U}(t,t_0)= 
\begin{pmatrix}
e^{\frac{1}{\hbar i}\int_{t_0}^{t}\left(-\sqrt{\frac{(E_{p1}(t')-E_{p2}(t'))^2}{4}+|t_{s12}(t')|^2}+\frac{E_{p1}(t')+E_{p2}(t')}{2}\right)dt'} & 0 \\
0 & e^{\frac{1}{\hbar i}\int_{t_0}^{t}\left(+\sqrt{\frac{(E_{p1}(t')-E_{p2}(t'))^2}{4}+|t_{s12}(t')|^2}+\frac{E_{p1}(t')+E_{p2}(t')}{2}\right)dt'} \\
\end{pmatrix},
\end{eqnarray}
\begin{eqnarray}
\ket{\psi_t}=c_{e1}(t_0)e^{\frac{1}{\hbar i}\int_{t_0}^{t}\left(-\sqrt{\frac{(E_{p1}(t')-E_{p2}(t'))^2}{4}+|t_{s12}(t')|^2}+\frac{E_{p1}(t')+E_{p2}(t')}{2}\right)dt'}\ket{E_1(t)}+ \nonumber \\ +c_{e2}(t_0)e^{\frac{1}{\hbar i}\int_{t_0}^{t}\left(+\sqrt{\frac{(E_{p1}(t')-E_{p2}(t'))^2}{4}+|t_{s12}(t')|^2}+\frac{E_{p1}(t')+E_{p2}(t')}{2}\right)dt'}\ket{E_2(t)}=, \nonumber \\
=c_{e1}(t_0)e^{\frac{1}{\hbar i}\int_{t_0}^{t}\left(-\sqrt{\frac{(E_{p1}(t')-E_{p2}(t'))^2}{4}+|t_{s12}(t')|^2}+\frac{E_{p1}(t')+E_{p2}(t')}{2}\right)dt'}
\begin{pmatrix}
\frac{((E_{p2}(t)-E_{p1}(t))+\sqrt{\frac{(E_{p2}(t)-E_{p1}(t))^2}{2}+|t_{s12}(t)|^2})e^{i phase(t_{s12}(t))}i)}{\sqrt{|t_s(t)|^2+((E_{p2}(t)-E_{p1}(t))+\sqrt{\frac{(E_{p2}(t)-E_{p1}(t))^2}{2}+|t_{s12}(t)|^2})^2}} \\
\frac{-|t_s(t)|)}{\sqrt{|t_s(t)|^2+((E_{p2}(t)-E_{p1}(t))+\sqrt{\frac{(E_{p2}(t)-E_{p1}(t))^2}{2}+|t_{s12}(t)|^2})^2}}\\
\end{pmatrix}_{x}+\nonumber \\
+c_{e2}(t_0)e^{\frac{1}{\hbar i}\int_{t_0}^{t}\left(+\sqrt{\frac{(E_{p1}(t')-E_{p2}(t'))^2}{4}+|t_{s12}(t')|^2}+\frac{E_{p1}(t')+E_{p2}(t')}{2}\right)dt'}
\begin{pmatrix}
\frac{(-(E_{p2}(t)-E_{p1}(t))+\sqrt{\frac{(E_{p2}(t)-E_{p1}(t))^2}{2}+|t_{s}(t)|^2})e^{-iphase(t_{s12}(t))}}{\sqrt{|t_s|^2+(-(E_{p2}(t)-E_{p1}(t))+\sqrt{\frac{(E_{p2}(t)-E_{p1}(t))^2}{2}+|t_{s12}(t)|^2})^2}} \\
\frac{+|t_s(t)|}{\sqrt{|t_s|^2+(-(E_{p2}(t)-E_{p1}(t))+\sqrt{\frac{(E_{p2}(t)-E_{p1}(t))^2}{2}+|t_{s12}(t)|^2})^2}} \\
\end{pmatrix}_{x}.\nonumber \\
\end{eqnarray}
Here, $c_{e1}(t_0)$ and $c_{e2}(t_0)$ describe the qubit in the energy representation at the initial time $t_0$, so $|c_{e1}(t_0)|^2+|c_{e2}(t_0)|^2=1$. Such presented evolution of position-based qubit is under the circumstances of small adiabatic changes in $t_s(t)$ and in
$E_{p1}(t)$, $E_{p2}(t)$. It is not the case of a qubit subjected to the rapid AC field that will support the existence of resonant states \cite{Pomorski_spie}.




\vspace{+10mm}

\section{Non-local realism in quantum mechanical picture}

Quantum mechanics merely provides a probabilistic description of physical processes, which does not support classical determinism but only stochastic one. A particle can be localized within a certain area of space when it is in a point-like potential minimum or it can be distributed over a large area as it is the case of conductive electron in metal. Once a measurement is conducted on the particle, its position can be determined exactly but at the price of that particle's momentum being highly perturbed, thus essentially losing its momentum information. In other words, one cannot fully determine both position and momentum of the particle. This is expressed in the non-commutation relation between the momentum and position and it leads to the Heisenberg principle.

In a very real sense, the quantum particle is like a classical particle under very large noise, so it is pointless to talk about the individual particle's position but it makes more sense to talk about the probability of finding the particle in a given ensemble of particles. We say that a canonical ensemble is attached to the individual particle's behavior. Thus, while dealing with a conglomerate of particles we are dealing with a statistical ensemble [of a single particle] attached to another statistical ensemble of environment in which the given particle is placed. Such reasoning indeed draws analogies of statistical mechanics with quantum mechanics. At some point, one can say that there is no significant difference between the quantum mechanical or classical particle under the impact of external potential.

Principle of locality holds for both classical and quantum pictures and two particles interact if they are close to each another. Coulomb electrostatic energy has the same formula in both the classical and quantum pictures. However, the first main difference is the fact that the quantum particle can be subjected to self-interference as it is the case of a double-slit experiment when the given wave (quantum particle) appears in certain regions with higher probabilities (higher wave intensity), while in other regions with lower probabilities. Self-interference requires that the wavefunction of a given particle is coherent which is strongly dependent on the environment. Self interference has a classical counterpart in the theory of waves as the given electromagnetic wave can interfere with itself.

There is however an effect that has no classical counterpart in the quantum picture: entanglement. It is the manifestation of a non-local correlation. In classical physics, it is however not surprising that when two particles are interacting, the change of state of one particle brings the change of state of the other. However, the surprising aspect is when the two particles are separated with essentially no interaction, the change of the state of one particle is affecting the state of another particles in an \emph{immediate} way. Such event is called ``spooky action at a distance" and is an example of the non-local correlation that can only occur in the quantum theory and is the manifestation of the particle entanglement.

In this work, we will describe the entanglement between a waveguide and position-based qubits as well as the entanglement between two far separated position-based qubits mediated by a waveguide. Most common picture of entanglement is illustrated by the Bell states. Bells states among the position-based qubit and quantum electromagnetic cavity are naturally given in this work
especially when we approach the limit of strong electromagnetic cavity and position dependent qubit interaction which is only partly achievable under the condition of qubit placed directly in the center of cavity. A first approach shall consider the perturbative interaction of resonant cavity and position-based qubit.

\begin{figure}
\centering
\includegraphics[width=6.0in]{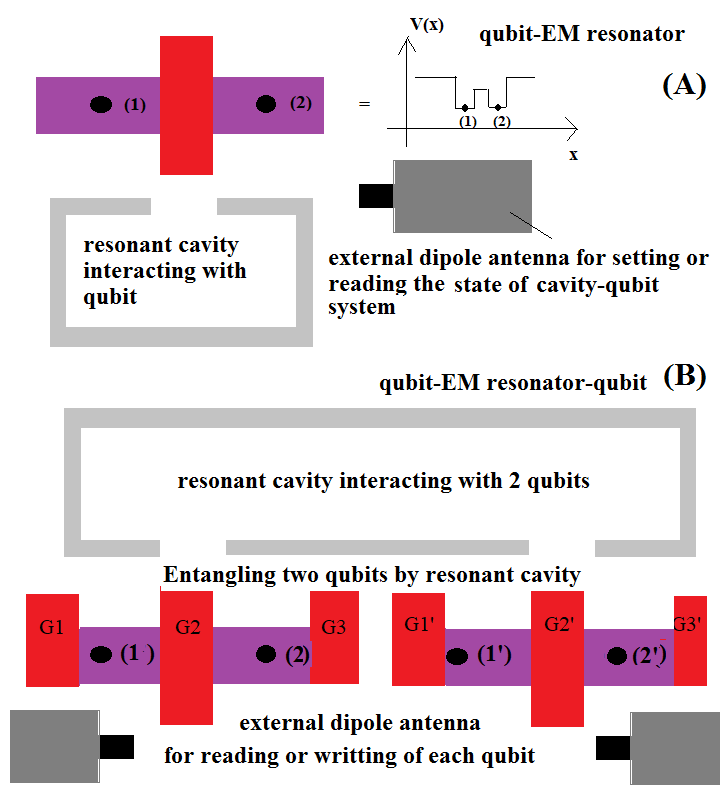} 
\caption{Position based qubit in RF field (A) and position based qubits placed at high distance interlinked by waveguide (B) \cite{Nbody}. Physical states of qubits are controlled by voltages applied to the gates G1, G2, G3 and G1', G2', G3'.}
\label{central2}
\end{figure}

\section{Interaction of electromagnetic cavity with position-dependent qubit}

We are now making a strong assumption that we are given an electromagnetic cavity (EC) that maintains quantum coherence. At the same time, we are dealing with a position-based semiconductor qubit that maintains its own quantum coherence, and that those two physical objects are interacting in a coherent way.
We are going to use Jaynes-Cumming Hamiltonian \cite{Jaynes} that describes the interaction
atom with cavity by means of electromagnetic field. In the simplest approach, the cavity Hamiltonian without dissipation (as it can be pre-assumed for the electromagnetic cavities with very high quality factor) is represented as
\begin{equation}
H_{cavity} = \hbar \omega_c (\frac{1}{2}+\hat{a}^{\dag}\hat{a}) = E_{\phi1} \ket{E_{\phi1}}\bra{E_{\phi1}} +E_{\phi2} \ket{E_{\phi2}}\bra{E_{\phi2}}+..=
\sum_{k=1}^{+\infty}\hbar \omega_c (\frac{1}{2}+k)\ket{E_{\phi_k}}\bra{E_{\phi_k}},
\end{equation}
where $\hat{a}^{\dag}$ ($\hat{a}$) is the photon creation (annihilation) operator and the number of photons in the cavity is given as $n=\hat{a}^{\dag}\hat{a}$. At the same time, we can represent the two-level qubit
system $H_{qubit} = E_g \ket{g}\bra{g} +E_e \ket{e}\bra{e}.$
The interaction Hamilonian is of the following form
$H_{qubit-cavity} = g(\sigma_{-}\hat{a}^{\dag}+\sigma_{+}\hat{a})$, 
where $\sigma_{-}=\sigma_{1}-i \sigma_{2}$, $\sigma_{+}=\sigma_{1}+i \sigma_{2}$ are expressed by Pauli matrices.
The qubity-cavity interaction has the electric-dipole nature, so quasiclassicaly we can write
\begin{eqnarray}
H_{qubit-cavity} = \hat{d}(t) \cdot \hat{E}(t) = g (\sigma_{-} + \sigma_{+})(\hat{a}+\hat{a}^{\dag}) \approx g(\sigma_{-}\hat{a}^{\dag}+\sigma_{+}\hat{a}).
\end{eqnarray}
It is worth mentioning that $H_{qubit-cavity}$ is time dependent since the electromagnetic field oscillates in the electromagnetic cavity. Also, an electron in the position-based qubits oscillates between two positions with its natural frequency that can be tuned by changing the height of potential barrier between neighboring quantum dots \cite{Nbody}.
We have neglected $g (\sigma_{-}\hat{a}+ \sigma_{+}\hat{a}^{\dag})$ and our approach is known as a rotating phase description of matter radiation interaction. Constant $g$ is depending on the distance between waveguide and position-dependent qubit as depicted in Fig.\,\ref{central2}.
During a photon emission from the qubit, the energy level is lowered. Reversely, during a photon absorption, the energy level of qubit is raised, which is seen in the term $\hat{a}\sigma_{+}$. The system Hamiltonian is given as
$H=H_{cavity}+H_{qubit}+H_{qubit-cavity}$.
It is not hard to construct the Hilbert space for Jaynes-Cumming Hamiltonian.
Essentially, we are considering the tensor product of qubit Hilbert space and electromagnetic cavity space.
\begin{eqnarray}
\ket{\psi}=\gamma_1\ket{E_{\phi_1}}\ket{E_g}+\gamma_2\ket{E_{\phi_1}}\ket{E_e}+\gamma_3\ket{E_{\phi_2}}\ket{E_g}+\gamma_4\ket{E_{\phi_2}}\ket{E_e} 
=
\begin{pmatrix}
\gamma_1  \\
\gamma_2  \\
\gamma_3  \\
\gamma_4  \\
\end{pmatrix},
1=\braket{\psi}{\psi} =|\gamma_1|^2+..+|\gamma_4|^2.
\end{eqnarray}
It implies that the probability of occupancy of $\ket{E_{\phi1}}$ state by qEQ is $|\gamma_1|^2+|\gamma_2|^2$, while the probability for qubit to be in the ground state is $|\gamma_1|^2+|\gamma_3|^2$ and the probability for finding q-state within the ground state of qubit and ground state of cavity is $|\gamma(1)|^2$.
Here, $\ket{g}$ and $\ket{e}$ stand for $E_g$ and $E_e$ energetic states of the position-based qubit, while $\ket{E_{\phi_1}}$ and $\ket{E_{\phi_2}}$ stand for cavity in the ground and excited states.
We have the following matrices $H_{qubit}+H_{cavity}$, $H_{qubit-cavity}$
\begin{eqnarray}
H_{qubit}+H_{cavity}=
\begin{pmatrix}
E_g+E_{\phi_1} & 0 & 0 & 0 \\
0 & E_e+E_{\phi_1} & 0 & 0 \\
0 & 0 & E_g+E_{\phi_2} & 0 \\
0 & 0 & 0 & E_e+E_{\phi_2} \\
\end{pmatrix} \nonumber \\ 
H_{qubit-cavity}=
\begin{pmatrix}
0 & 0 & 0 & 0 \\
0 & 0 & g_1 & 0 \\
0 & g_1^{*} & 0 & 0 \\
0 & 0 & 0 & 0 \\
\end{pmatrix} =g_1(\ket{E_{\phi_1}, E_e}\bra{E_{\phi_2}, E_g}+\ket{E_{\phi_2}, E_g}\bra{E_{\phi_1}, E_e}),
\end{eqnarray}
which implies
\begin{eqnarray}
H_{qubit}+H_{cavity}+H_{qubit-cavity}=
\begin{pmatrix}
E_g+E_{\phi_1} & 0 & 0 & 0 \\
0 & E_e+E_{\phi_1} & g_1(t) & 0 \\
0 & g_1(t)^{*} & E_g+E_{\phi_2} & 0 \\
0 & 0 & 0 & E_e+E_{\phi_2} \\
\end{pmatrix}.
\label{14}
\end{eqnarray}
The last Hamiltonian gives four eigenstates and one has two non-entangled states 
$\ket{E_1}=
\begin{pmatrix}
1 \\
0 \\
0 \\
0 \\
\end{pmatrix}=\ket{E_{\phi_1}}\ket{E_g}$,  
\newline
$\ket{E_2}=
\begin{pmatrix}
0 \\
0 \\
0 \\
1 \\
\end{pmatrix}=\ket{E_{\phi_2}}\ket{E_e}$, 
and two entangled states $\ket{E_3}, \ket{E_4}$ exist due to the non-zero coefficient $g_2$ and are given as 
\begin{eqnarray}
\nonumber \\
\ket{E_3}=
\begin{pmatrix}
0 \\
\frac{(E_e-E_g)-(E_{\phi_2}-E_{\phi_1})-\sqrt{((E_e-E_g)-(E_{\phi_2}-E_{\phi_1}))^2+4|g_1|^2}}{2g_1} \\
1 \\
0 \\
\end{pmatrix} \nonumber \\
=\Bigg[ \frac{(E_e-E_g)-(E_{\phi_2}-E_{\phi_1})}{2g_1}
-\frac{\sqrt{((E_e-E_g)-(E_{\phi_2}-E_{\phi_1}))^2+4|g_1|^2} }{2g_1} \Bigg] 
\ket{E_{\phi_1}}\ket{E_e}
+ \ket{E_{\phi_2}}\ket{E_g}, \nonumber \\
\ket{E_4}=
\begin{pmatrix}
0 \\
\frac{(E_e-E_g)-(E_{ph_2}-E_{ph_1})+\sqrt{((E_e-E_g)-(E_{\phi_2}-E_{\phi_1}))^2+4|g_1|^2}}{2g_1} \\
1 \\
0 \\
\end{pmatrix},
\end{eqnarray}
\normalsize
One obtains four corresponding eigenenergies of the form
$E_1=E_g+E_{\phi_1}$, 
$E_2=E_e+E_{\phi_2}$ and  
\begin{eqnarray}
 E_3=\frac{1}{2}(E_g+E_e+E_{\phi_1}+E_{\phi_2} 
 -\sqrt{((E_e-E_g)-(E_{\phi_2}-E_{\phi_1}))^2+4|g_1|^2}, \nonumber \\
 E_4=\frac{1}{2}(E_g+E_e+E_{\phi_1}+E_{\phi_2} 
 +\sqrt{((E_e-E_g)-(E_{\phi_2}-E_{\phi_1}))^2+4|g_1|^2}, 
\end{eqnarray}
It shall be underlined that $g_2(t)=\lambda_0 E_{0,x1}$, where $E_{0,x1}$ is intensity of 1-st EM mode 
and basically $g_2$ is very small in comparison with $E_{\phi1},E_{\phi2}, E_{g}, E_{e} $ energies.
In all cases, $E_2>E_1$. We recognize that the states corresponding to eigenenergies $E_3$ and $E_4$ are entangled states of matter and radiation while the states corresponding to eigenenergies $E_1$ and $E_2$ are non-entangled states of matter and radiation.
In particular, if state $E_3$ is subjected to the measurement of a number of photons and value 1 was encountered, then it implies that the position-based qubit is in the excited state corresponding to energy $E_e$. Otherwise, if the number of photons encountered is 2, then the state of qubit is $E_g$.
It is quite easy to determine the quantum evolution under the action of Hamiltonian (\ref{14}). We have
$\ket{\psi_t}=e^{\int_{t_0}^{t}\frac{1}{\hbar i}\hat{H}(t')dt'}\ket{\psi_{t_0}}=\hat{U}(t,t_0)\ket{\psi_{t_0}},$
where elements of evolution operator $\hat{U}(t,t_0)$ (expressed as 4$\times$4 matrix in this case) are given in the analytical way
\begin{eqnarray}
\hat{U}(t,t_0)_{1,2}=\hat{U}(t,t_0)_{1,3}=\hat{U}(t,t_0)_{1,4}=\hat{U}(t,t_0)_{4,2}=\hat{U}(t,t_0)_{4,3}=\hat{U}(t,t_0)_{2,1}=\hat{U}(t,t_0)_{3,1}=0, \nonumber \\
\hat{U}(t,t_0)_{1,1}=e^{\frac{1}{\hbar i}((\int_{t_0}^{t}dt'E_g(t'))+(t-t_0)E_{\phi1})}, 
\hat{U}(t,t_0)_{4,4}=e^{\frac{1}{\hbar i}((\int_{t_0}^{t}dt'E_e(t'))+(t-t_0)E_{\phi2})}, \nonumber \\
\end{eqnarray}
\begin{eqnarray}
\hat{U}(t,t_0)_{2,2}= 
\frac{e^{-\frac{i \left(\sqrt{(\int_{t_0}^{t}(E_e(t')+E_{\phi_1}-E_g(t')-E_{\phi_2})dt')^2+\left(|2G(t)|^2\right)}+(\int_{t_0}^{t}E_e(t')+E_{\phi_1}+E_g(t')+E_{\phi_2})dt')\right)}{2\hbar}}}{2 \sqrt{(\int_{t_0}^{t}(E_e(t')+E_{\phi_1}-E_g(t')-E_{\phi_2})dt')^2+4 \left(|G(t)|^2\right)}} \nonumber \\
\times \Bigg[ -\left( \int_{t_0}^{t}dt'E_e(t'))+(t-t_0)E_{\phi1} \right) \times 
\left(-e^{\frac{i \sqrt{(\int_{t_0}^{t}(E_e(t')+E_{\phi_1}-E_g(t')-E_{\phi_2})dt')^2+4
   \left(|G(t)|^2\right)}}{\hbar}}\right)+ \nonumber \\ +\sqrt{(\int_{t_0}^{t}(E_e(t')+E_{\phi_1}-E_g(t')-E_{\phi_2})dt')^2+4
   \left(|G(t)|^2\right)}+\nonumber \\ +(\int_{t_0}^{t}(E_e(t')+E_{\phi_1}-E_g(t')-E_{\phi_2})dt')+ 
+\left(e^{\frac{i \sqrt{(\int_{t_0}^{t}(E_e(t')+E_{\phi_1}-E_g(t')-E_{\phi_2})dt')^2+4
   \left(|G(t)|^2\right)}}{\hbar}}\right)\times \nonumber \\ \times \Big[(\int_{t_0}^{t}dt' E_g(t')+E_{\phi2}(t-t_0)) 
+\sqrt{(\int_{t_0}^{t}(E_e(t')+E_{\phi_1}-E_g(t')-E_{\phi_2})dt')^2+\left(|2G(t)|^2\right)}\Big] \Bigg], 
\end{eqnarray}
\begin{eqnarray}
\hat{U}(t,t_0)_{3,3}= 
\frac{e^{-\frac{i \left(\sqrt{(\int_{t_0}^{t}(E_e(t')+E_{\phi_1}-E_g(t')-E_{\phi_2})dt')^2+\left(|2G(t)|^2\right)}+(\int_{t_0}^{t}E_e(t')+E_{\phi_1}+E_g(t')+E_{\phi_2})dt')\right)}{2\hbar}}}{2 \sqrt{(\int_{t_0}^{t}(E_e(t')+E_{\phi_1}-E_g(t')-E_{\phi_2})dt')^2+4 \left(|G(t)|^2\right)}}
\Bigg[ \left( \int_{t_0}^{t}dt'E_e(t'))+(t-t_0)E_{\phi1} \right) \times \nonumber \\ \times \left(-e^{\frac{i \sqrt{(\int_{t_0}^{t}(E_e(t')+E_{\phi_1}-E_g(t')-E_{\phi_2})dt')^2+4
   \left(|G(t)|^2\right)}}{\hbar}}\right)  
+\sqrt{(\int_{t_0}^{t}(E_e(t')+E_{\phi_1}-E_g(t')-E_{\phi_2})dt')^2+4
   \left(|G(t)|^2\right)}+\nonumber \\ -(\int_{t_0}^{t}(E_e(t')+E_{\phi_1}-E_g(t')-E_{\phi_2})dt') 
+\left(e^{\frac{i \sqrt{(\int_{t_0}^{t}(E_e(t')+E_{\phi_1}-E_g(t')-E_{\phi_2})dt')^2+4
   \left(|G(t)|^2\right)}}{\hbar}}\right) 
\Big[-(\int_{t_0}^{t}dt' E_g(t')+E_{\phi2}(t-t_0))+ \nonumber \\ +\sqrt{(\int_{t_0}^{t}(E_e(t')+E_{\phi_1}-E_g(t')-E_{\phi_2})dt')^2+\left(|2G(t)|^2\right)}\Big] \Bigg].  \nonumber \\
\end{eqnarray}

We also have
\begin{eqnarray}
\hat{U}(t,t_0)_{3,2} =-2 i (G^{*}(t)e^{-\frac{i ( (\int_{t_0}^{t}(\Delta_q(t')+2E_{g}(t'))dt')+(2E_{\phi1}+\Delta_{EC})(t-t_0))}{2 \hbar}}) \frac{ \sin \left(\frac{\sqrt{\Big[(( (\int_{t_0}^{t}(\Delta_q(t'))dt')-(\Delta_{EC})(t-t_0)))\Big]^2+4
   \left(|G(t)|^2\right)}}{2 \hbar}\right)}{\Bigg[\frac{\sqrt{(\Big[ (\int_{t_0}^{t}(\Delta_{q}(t')dt')-(\Delta_{EC})(t-t_0))\Big]^2+4
   \left(|G(t)|^2\right)}}{2\hbar}\Bigg] 2\hbar }
\end{eqnarray}
where we have introduced the energy level differences between the first excited and ground states of isolated qubit $\Delta_q(t)=E_e(t)-E_g(t)$, and between the first excited and ground states of quantum electromagnetic cavity $\Delta_{EC}(t)=E_{\phi2}(t)-E_{\phi1}(t)$, $G(t)=\int_{t_0}^{t}dt'g(t')$. In future considerations, it will be helpful to operate with functions $EI_g(t)=\int_{t_0}^{t}dt'E_g(t'), EI_e(t)=\int_{t_0}^{t}dt'E_e(t'), EI_{\phi1(2)}(t)=\int_{t_0}^{t}dt'E_{\phi1(2)}(t')$.
It is worth underlining that we can relatively easily tune $E_e(t)$ and $E_g(t)$ with time by changing voltages controlling the position-based qubits. However, it is not the case with the quantum EC where varying $E_{\phi1}$ and $E_{\phi2}$ with time is rather technologically difficult so we pre-assume that $E_{\phi1}$ and $E_{\phi2}$ are time invariant. This is because EC eigenenergies depend on the geometry, which is fixed, and incorporated in the qEC nanostructure. Under certain circumstances, this geometry could be changed by mechanical pressure or in some electrical manner. We can renormalize the operator $\hat{U}(t,t_0)$, so 
$\hat{U}(t,t_0)_n=\frac{1}{e^{\frac{1}{\hbar i}\int_{t_0}^{t}dt'(E_g(t')+E_e(t')+E_{\phi1}(t')+E_{\phi1}(t'))}}\hat{U}(t,t_0)$
and in such case $\det(\hat{U}(t,t_0)_n)=1$.
 We recognize that
\begin{eqnarray}
\gamma_1(t)=\hat{U}(t,t_0)_{1,2}\gamma_1(t_0),
\gamma_2(t)=\hat{U}(t,t_0)_{2,2}\gamma_2(t_0)+\hat{U}(t,t_0)_{2,3}\gamma_3(t_0), 
\gamma_3(t)=\hat{U}(t,t_0)_{3,2}\gamma_2(t_0)+\hat{U}(t,t_0)_{3,3}\gamma_3(t_0)
\end{eqnarray}
which implies that for the time dependent eigenenergies of qubit and cavity, $|\gamma_1(t)|=|\gamma_1(t_0)|=\rm const_1$, $|\gamma_4(t)|=|\gamma_4(t_0)=\rm const_2|$ (as is the case of non-interacting qubit with electromagnetic cavity so it is the case here of time-independent Hamiltonian for both separated physical systems). Probability for having the qubit in the excited state is
\begin{eqnarray}
p_{QC: excited}(t)=|\gamma_2(t)|^2+|\gamma_4(t)|^2
=|\gamma_4(t_0)|^2+|\hat{U}(t,t_0)_{2,2}\gamma_2(t_0)+\hat{U}(t,t_0)_{2,3}\gamma_3(t_0)|^2
\end{eqnarray}
and the probability for having the cavity in the excited state is
\begin{eqnarray}
p_{qEC:Excited}(t)=|\gamma_3(t)|^2+|\gamma_4(t)|^2 
=|\hat{U}(t,t_0)_{3,2}\gamma_2(t_0)+\hat{U}(t,t_0)_{3,3}\gamma_3(t_0)|^2+|\gamma_4(t_0)|^2.
\end{eqnarray}
In order to trace $p_{QC: excited}(t)$ and $p_{qEC: excited}(t)$ with time, one needs six initial parameters plus knowledge of two cavity eigenenergies plus full knowledge of four functions with time $E_{p1}(t), E_{p2}(t), |t_{s12}(t)|, {\rm phase}(t_{s12}(t))$. The last four functions can be directly translated to $E_{e}(t), E_g(t)$ as well as initial qubit eigenenergy states.
It shall be underlined that $|\hat{U}(t,t_0)_{3,2}|^2$ describes the mutual energy flow between the qubit's ground state into excited state (the qubit's ground state becomes more populated and consequently the qubit's exited state becomes less populated) and population of excited state of electromagnetic cavity with depopulation of lower cavity eigenenergy state. It is therefore instructive to compute $|\hat{U}(t,t_0)_{3,2}|^2$| with time in a simple case, as when, for example, all eigenenergies $E_g, E_e, E_{\phi1}, E_{\phi2}$ are time independent. Let us assume that the qubit is in its excited state and that the cavity is in the state of eigenenergy $E_{\phi1}$. It implies that there exists a standing electromagnetic wave in the cavity at qubit position $x_q$ given as
$E_{x_q}(t)=E_{x,o1}(x_q)\sin(E_{\phi1}t+p_0)$ that leads to the expression $|G(t)|=|g(E_{x,o1}(x_q)|t=constant \times t$. Thus, we have $|\hat{U}(t,t_0)_{3,2}|^2$ given as
\vspace{-2mm}
\begin{eqnarray}
|\hat{U}(t,t_0)_{3,2}|^2= \Bigg[4|G(t)^2|\frac{ \sin \left(\frac{\sqrt{\Big[(( (\int_{t_0}^{t}(\Delta_q(t'))dt')-(\Delta_{EC})(t-t_0)))\Big]^2+4
   \left(|G(t)|^2\right)}}{2 \hbar}\right)}{\sqrt{(\Big[ (\int_{t_0}^{t}(\Delta_{q}(t')dt')-(\Delta_{EC})(t-t_0))\Bigg]^2+4
   \left(|G(t)|^2\right)}}\Bigg]^2 \nonumber \\
=s_04|(|g(E_{x,o1}(x_q))|^2(t-t_0)^2) \times \nonumber
 \Bigg[\frac{ \sin \left(\frac{\sqrt{\Big[(\int_{t_0}^{t}\Delta_q(t')dt')-\Delta_{EC}(t-t_0)\Big]^2+4|g(E_{x,o1}(x_q))|^2t^2
   }}{2 \hbar}\right)}{\frac{\sqrt{[(\int_{t_0}^{t}\Delta_{q}(t')dt')-\Delta_{EC}(t-t_0)]^2+4|g(E_{x,o1}(x_q))|^2t^2
   }}{2\hbar}}\Bigg]^2. 
\end{eqnarray}

\begin{figure}
\centering
\includegraphics[scale=0.95]{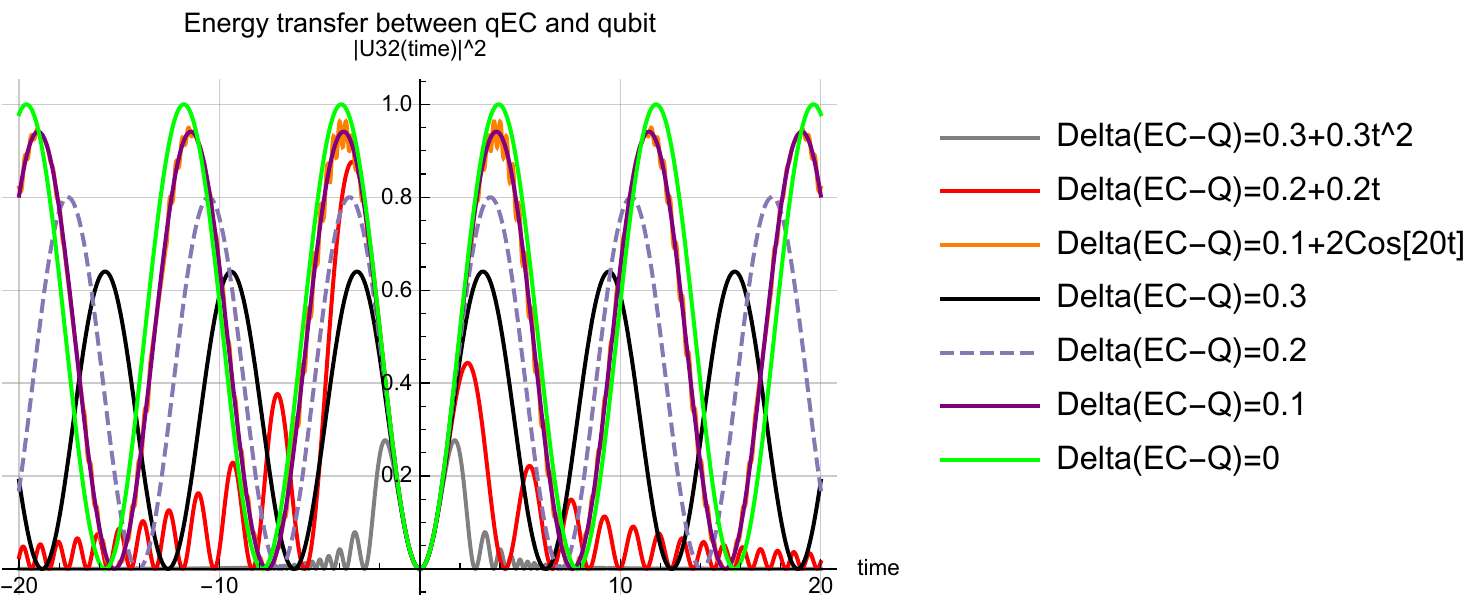}
\caption{Example of energy transfer coefficient vs. time for different energy differences $(E_e(t)-E_g(t))-(E_{\phi1}-E_{\phi2}))=\Delta_{q}(t)-\Delta_{EC}=\Delta(EC-Q)$. The ability of changing with time qubit eigenergy states gives the possibility of engineering of energy transfer coefficient with time.   }
\label{fig:3}
\end{figure}

We can shape this function by engineering dependence of $(\Delta_{q}(t)-\Delta_{EC})$. Since we can tune the energy difference between the excited and ground states in the position-dependent qubit by setting proper voltages across the controlling gates it is possible to obtain family of functions $\Delta_{q}(t)$. Various shapes of energy transfer function are depicted in Fig.3. In particular we can set $(\Delta_{q}-\Delta_{EC})=0$, what implies $E_e-E_g=E_{\phi2}-E_{\phi1}$ so we can obtain a simplified version of energy transfer function $|\hat{U}(t,t_0)_{3,2}|^2$ that has the form
\vspace{-2mm}
\begin{eqnarray}
|\hat{U}(t,t_0)_{3,2}|^2
 =s_1 \left( \sin(\frac{|g(E_{x,o1}(x_q)|}{\hbar}|t\right)^2=s_1 \sin(const_2 t)^2,
\end{eqnarray}
so we have only one oscillation frequency for energy transfer function.
It is instructive to see the last two presented energy transfer functions on one plot, as depicted in Fig.\,\ref{fig:3} for parameters $\hbar=1$, $E_{\phi1}=1$, $E_{x,o1}=1$ (Scenario 1-7)  when we set $\Delta_q-\Delta_{EC}=0$, $0.1$, $0.2$, $0.3$, $0.1+2\cos(20t)$, $0.2+0.2t$, $0.3+0.3t^2$ ). 
Specified scenarios are justified technologically when cavity is much bigger than position based qubit and physically under an assumption of weak electric fields. Clearly it is visible that the system of position based qubit interacting with quantum resonator provides rich tunnable family of functions $|\hat{U}(t,t_0)_{3,2}|^2$.
\subsection{Case of 2 qubits interacting via a waveguide over a distance and teleportation}

We have the following Hamiltonian for two qubits interacting with a waveguide in the case when qubit 1 is relatively far from qubit 2. If the waveguide has length $L$ and $c$ is the speed of signal propagating along the waveguide, we have $\Delta t=L/c$ and Hamiltonian is of the form:
\begin{eqnarray}
H=(E_{\phi1}\ket{E_{\phi_1}}\bra{E_{\phi_1}}+E_{\phi2}\ket{\phi_2}\bra{E_{\phi_2}})I_{qubit1}I_{qubit2} + \nonumber \\
+I_{cavity}(E_{g1}\ket{g_1}\bra{g_1}+E_{e1}\ket{e_1}\bra{e_1})I_{qubit2} 
 + I_{cavity}I_{qubit1}(E_{g2}\ket{g_2}\ket{g_2}+E_{e2}\ket{e_2}\bra{e_2}) + \nonumber \\ +g_1f_1(t)[(\ket{E_{\phi_1}}\bra{E_{\phi_2}})(\ket{e_1}\bra{g_1})
 +(\ket{E_{\phi_2}}\bra{E_{\phi_1}})(\ket{g_1}\bra{e_1}]I_{qubit2}+ \nonumber \\ +g_2f_1(t+\Delta t)[(\ket{\phi_1}\ket{\phi_2})I_{qubit1}(\ket{e_2}\bra{g_2})
 +(\ket{E_{\phi_2}}\bra{E_{\phi_1}} )I_{qubit1}(\ket{g_2}\bra{e_2})].
\end{eqnarray}
It is formally a system of three interating quantum bodies (qubit1)-(resonator or waveguide)-(qubit2), in which qubit~1 cannot directly interact with qubit 2 and so the quantum state has the form
\begin{eqnarray}
\ket{\psi(t)}=\alpha_1(t)\ket{E_{\phi_1}}\ket{g_1}\ket{g_2}+\alpha_2(t)\ket{E_{\phi_1}}\ket{g_1}\ket{e_2}
+\alpha_3(t)\ket{E_{\phi_1}}\ket{e_1}\ket{g_2}+ \nonumber \\
+\alpha_4(t)\ket{E_{\phi_1}}\ket{e_1}\ket{e_2}+ 
\alpha_5(t)\ket{E_{\phi_2}}\ket{g_1}\ket{g_2}+\alpha_6(t)\ket{E_{\phi_2}}\ket{g_1}\ket{e_2}+ 
\alpha_7(t)\ket{E_{\phi_2}}\ket{e_1}\ket{g_2}+\alpha_8(t)\ket{E_{\phi_2}}\ket{e_1}\ket{e_2}
\end{eqnarray}
There is a presence of two functions, $f(t)$ and $f(t+\Delta t)$. Here, $\Delta t$ is dependent on the distance between the two holes in electromagnetic cavity depicted in Fig\,\ref{central2}(B).
The normalization condition is fulfilled by $|\alpha_1(t)|^2+..|\alpha_8(t)|^2=1$. The system Hamiltonian matrix $\hat{H}_s$ is of the structure as given below
\small
\begin{eqnarray*}
\hat{H}_s=\nonumber \\
\begin{pmatrix}
E_{g1}+E_{g2}+E_{\phi1} & 0 & 0 & 0 & 0 & 0 & 0 & 0 \\
0                                          & E_{g1}+E_{e2}+E_{\phi1}  & 0 & 0 & f_1(t)e^{-i d_2 t} g_2  & 0 & 0 & 0 \\
0                                         & 0                                             & E_{e1}+E_{g2}+E_{\phi1} & 0 & f_1(t)g_1e^{-i d_1 t}  & 0 & 0 & 0 \\
0 & 0 & 0 & E_{e1}+E_{e2}+E_{\phi1} & 0 & g_1 f_1(t)e^{-i d_1 t} & g_2f_1(t)e^{-i d_2 t} & 0 \\
0 & f_1(t)e^{i d_2 t} g_2 & f_1 (t)e^{i d_1 t} g_1 & 0 & E_{g1}+E_{g2}+E_{\phi2} & 0 & 0 & 0 \\
0 & 0 & 0 & g_1 f_1(t)e^{i d_1 t} & 0 & E_{g1}+E_{e2}+E_{\phi2} & 0 & 0 \\
0 & 0 & 0 & f_1(t)e^{i d_2 t} g_2 & 0 & 0 & E_{e1}+E_{g2}+E_{\phi2} & 0 \\
0 & 0 & 0 & 0 & 0 & 0 & 0 & E_{e1}+E_{e2}+E_{\phi2} \\
\end{pmatrix}
\end{eqnarray*}
\normalsize

This matrix can be simplified. We can pre-assume that $g_1f_1(t)=g f(t)e^{id_1(t)}$ and $g_2f_2(t)=g f(t)e^{id_2(t)}$ 
and we can divide all matrix entries by this value. The second simplification is by $E_g=E_{g1}=E_{g2}=E_{\phi1}=E_{\phi2}-E_{\phi1}=E_{e1}-E_{g1}=E_{e2}-E_{g2}$. In such a case, we obtain simplified Hamiltonian as
\begin{eqnarray}\hat{H}=
\begin{pmatrix}
3E_g & 0    & 0    & 0 & 0  & 0 & 0 & 0 \\
0    & 4E_g & 0    & 0 & g_2 e^{-i d_2 (t)}  & 0 & 0 & 0 \\
0    & 0    & 4E_g & 0 & g_1 e^{-i d_1 (t)}  & 0 & 0 & 0 \\
0    & 0    & 0    & 5E_g & 0 & g_1 e^{-id_1t}  & g_2 e^{-id_2(t)} & 0  \\
0 & g_2 e^{i d_2 (t)} & g_1 e^{i d_1 (t)}  &  0 & 4E_g & 0 & 0 & 0 \\
0 & 0 & 0 & g_1 e^{i d_1 (t)} & 0  & 5E_g & 0 & 0 \\
0 & 0 & 0 & g_2 e^{i d_2 (t)} & 0  & 0 & 5E_g & 0 \\
0 & 0 & 0 & 0 & 0  & 0 & 0 & 6E_g \\
\end{pmatrix}.
\end{eqnarray}
It shall be underlined that $g_1$ and $g_2$ are proportional to the electric field in the resonator's cavity that is time and space dependent. The exact determination of those coefficient requires the canonical quantization of EM field \cite{Birula}.
If we are dealing with two or more qubits, we assume that they are are individually coupled to the EM field. It can mean that qubits placed at different geometrical place in cavity or in its proximity catch different EM modes in different way since amplitude of EM standing field depends on the geometrical position. Even more complicated is the case of wave-guide where we can have "traveling EM" waves. EM in wave-guide or in cavity oscillates with its own frequency. In wave-guide it is obvious that very distance qubits will pick-up electromagnetic signal at different phase.  Ability for distanced qubit to catch EM wave at different phases is expressed by phase factors $e^{i d_1 (t)}$ and $e^{i d_2 (t)}$.
Different values of $d_1(t)$ and $d_2(t)$ are dependent on the distance between holes in the electromagnetic activity depicted in Fig.\,\ref{central2}(B). 
The last Hamiltonian matrix has the following energy eigenvalues: $3E_g,4E_g,5E_g,6E_g,4E_g-\sqrt{g_1^2+g_2^2},5E_g-\sqrt{g_1^2+g_2^2}$, $4E_g+\sqrt{g_1^2+g_2^2}$, $5E_g+\sqrt{g_1^2+g_2^2}$.

In a general case, $g_1$ depends on how a waveguide with a hole is close to the position-dependent qubit. Otherwise, the position-dependent qubit must be placed in the resonant cavity.
We denote the normalized states with $n$, so $\ket{E_k}_n=\frac{\ket{E_k}}{N_k}$ is the normalized state $\ket{E_k}$. All eight energy eigenstates are given below
\begin{eqnarray}
\ket{E_1}=
\begin{pmatrix}
1 \\
0 \\
0 \\
0 \\
0 \\
0 \\
0 \\
0 \\
\end{pmatrix}=\ket{E_{\phi_1}}\ket{g_1}\ket{g_2}=\ket{E_1}_n
=\frac{1}{2}\ket{E_{\phi_1}}(\ket{x_1}-\ket{x_2})(\ket{x_{1'}}-\ket{x_{2'}}), N_1=1, \frac{\ket{E_1}}{N_1}=\ket{E_1}_n,
\end{eqnarray}
 and with the assumption $E_p=E_{p1}=E_{p2}=E_{p1'}=E_{p2'}$, we obtain
\begin{eqnarray}
\ket{E_2}=
\begin{pmatrix}
0 \\
-(g_1/g_2)e^{i(-d_2+d_1)} \\
+1 \\
0 \\
0 \\
0 \\
0 \\
0 \\
\end{pmatrix}=
-(\frac{g_1}{g_2}e^{i(-d_2+d_1)})\ket{E_{\phi_1}}\ket{g_1}\ket{e_2}+\ket{E_{\phi_1}}\ket{e_1}\ket{g_2}, 
N_2=\frac{1}{\sqrt{1+|-(g_1/g_2)e^{i(-d_2+d_1)}|^2}}, \nonumber \\
\ket{E_3}= 
\begin{pmatrix}
0 \\
0 \\
0 \\
0 \\
0 \\
-\frac{g_2}{g_1}e^{i(-d_2+d_1)} \\
1 \\
0 \\
\end{pmatrix}=
-\frac{g_2}{g_1}e^{i(-d_2+d_1)}\ket{E_{\phi_2}}\ket{g_1}\ket{e_2}+\ket{E_{\phi_2}}\ket{e_1}\ket{g_2}, 
N_3=\frac{1}{\sqrt{1+|-\frac{g_2}{g_1}e^{i(-d_2+d_1)}|^2}}, \nonumber \\
\ket{E_4}=\ket{E_4}_n=
\begin{pmatrix}
0 \\
0 \\
0 \\
0 \\
0 \\
0 \\
0 \\
1 \\
\end{pmatrix}=+\ket{E_{\phi_2}}\ket{e_1}\ket{e_2}, N_4=1, 
N_5=\frac{1}{\sqrt{1+|-\frac{g_2e^{id_2}}{\sqrt{g_1^2+g_2^2}}|^2+|-\frac{g_1e^{id_1}}{\sqrt{g_1^2+g_2^2}} |^2}}, \nonumber \\
\ket{E_5}=
\begin{pmatrix}
0 \\
-\frac{g_2e^{id_2}}{\sqrt{g_1^2+g_2^2}} \\
-\frac{g_1e^{id_1}}{\sqrt{g_1^2+g_2^2}} \\
0 \\
1 \\
0 \\
0 \\
0 \\
\end{pmatrix}=-\frac{g_2e^{id_2}}{\sqrt{g_1^2+g_2^2}}\ket{E_{\phi_1}}\ket{g_1}\ket{e_2}
- \frac{g_1e^{id_1}}{\sqrt{g_1^2+g_2^2}}\ket{E_{\phi_1}}\ket{e_1}\ket{g_2}
+\frac{1}{\sqrt{2}}\ket{E_{\phi_1}}\ket{e_1}\ket{e_2}, \nonumber \\
\ket{E_6}=
\begin{pmatrix}
0 \\
0 \\
0 \\
-\frac{e^{2id_2}\sqrt{g_1^2+g_2^2}}{g_2} \\
0 \\
1 \\
1 \\
0 \\
\end{pmatrix}= 
 -\frac{e^{2id_2}\sqrt{g_1^2+g_2^2}}{g_2}\ket{E_{\phi_1}}\ket{e_1}\ket{e_2}+
\ket{E_{\phi_2}}\ket{e_1}\ket{g_2}+\ket{E_{\phi_2}}\ket{g_1}\ket{e_2} , 
N_6=\frac{1}{2+|-\frac{e^{2id_2}\sqrt{g_1^2+g_2^2}}{g_2}|^2}, \nonumber \\
\ket{E_7}= 
\begin{pmatrix}
0 \\
e^{-id_2}\frac{g_2}{\sqrt{g_1^2+g_2^2}} \\
e^{-id_1}\frac{g_1}{\sqrt{g_1^2+g_2^2}} \\
0 \\
1 \\
0 \\
0 \\
0 \\
\end{pmatrix} 
=e^{-id_2}\frac{g_2}{\sqrt{g_1^2+g_2^2}}(\ket{E_{\phi_1}}\ket{e_1}\ket{g_2}+
e^{-id_1}\frac{g_1}{\sqrt{g_1^2+g_2^2}}(\ket{E_{\phi_1}}\ket{g_1}\ket{e_2}+
\ket{E_{\phi_2}}\ket{g_1}\ket{g_2}, \nonumber \\
N_7=\frac{1}{\sqrt{1+|e^{-id_2}\frac{g_2}{\sqrt{g_1^2+g_2^2}}|^2+|e^{-id_1}\frac{g_1}{\sqrt{g_1^2+g_2^2}}|^2}}, 
 N_8=\frac{1}{\sqrt{1+|e^{-i(2d_1+d_2)}\frac{\sqrt{g_1^2+g_2^2}}{g_2}|^2+|e^{-id_2+id_1}\frac{g_1}{g_2}|^2}}, \nonumber \\
\end{eqnarray}
\begin{eqnarray}
\ket{E_8} =
\begin{pmatrix}
0 \\
0 \\
0 \\
e^{-i(2d_1+d_2)}\frac{\sqrt{g_1^2+g_2^2}}{g_2} \\
0 \\
e^{-id_2+id_1}\frac{g_1}{g_2} \\
1 \\
0 \\
\end{pmatrix} =
e^{-i(2d_1+d_2)}\frac{\sqrt{g_1^2+g_2^2}}{g_2}\ket{E_{\phi_1}}\ket{e_1}\ket{e_2}+ 
e^{-id_2+id_1}\frac{g_1}{g_2}\ket{E_{\phi_2}}\ket{e_1}\ket{g_2}+\ket{E_{\phi_2}}\ket{g_1}\ket{e_2}, \nonumber \\
\end{eqnarray}
We observe that six eigenstates among the eight eigenstates (except for the $E_1$ and $E_4$ eigenstates that are not entangled) are entangled in the energy basis.
It is noticeable to underline that all eight energy eigenstates are entangled in the position-based representation, which was pointed out for the case of all $E_p$ values corresponding to the nodes in two different qubits.
The quantum state dynamics of the system $QC_1-EC-QC_2$ can be written as
\begin{equation}
\ket{\psi(t)}=c_{e1} e^{\frac{\hbar}{i}\int_{t_0}^{t}E_1(t')dt'}\ket{E_1}_n+..+c_{e8}e^{\frac{\hbar}{i}\int_{t_0}^t dt' E_8(t')}\ket{E_8}_n,.
\end{equation}
where the normalization relation takes place $|c_{e1}|^2+..+|c_{e8}|^2=1$,
in which the eigenenergy states are orthonormal since $\bra{E_k} \ket{E_s} = \delta_{k,s}$.

\section{Essence of quantum communication between qubits entangled by electromagnetic cavity over long distance}

We have observed the entanglement between the neighboring qubit and electromagnetic cavity (EC). This entanglement between qubit 1 and EC as well as entanglement between qubit 2 and EC will lead to the entanglement between qubit 1 and qubit 2.
Such reasoning can be generalized for the case of $N$ qubits entangled to one electromagnetic cavity (EC).
Let us exercise a possible communication scheme by the use of projectors.
Let us enforce qubit 1 to be in the excited state, which can be achieved by the use of one antenna (as the left antenna in Fig.\,\ref{central2}). It requires a delivery of a certain electromagnetic pulse by antenna 1. This pulse will also bring some secondary effect on qEC as well on another distant qubit. However, due to the simplifications in the conducted analysis we will neglect the secondary effects.
In such a case, the projector becomes
$\hat{P}_{e1}=\hat{I}_{cavity}\times \ket{e_1}\bra{e_1} \times \hat{I}_{qubit2}= 
(\ket{E_{\phi1}}\bra{E_{\phi1}}+\ket{E_{\phi2}}\bra{E_{\phi2}})\ket{e_1}\bra{e_1} (\ket{g_{2}}\bra{g_{2}}+\ket{e_{2}}\bra{e_{2}})$.
Let us exercise the action of $\hat{P}_{e1}$ operator on quantum state $\ket{\psi}=\ket{E_2}=\frac{1}{N_2}(-(\frac{g_1}{g_2}e^{i(-d_2+d_1)}\ket{E_{\phi_1}}\ket{g_1}\ket{e_2})+\ket{E_{\phi_1}}\ket{e_1}\ket{g_2})$ so we obtain the quntum state after bringing the first qubit into the excited state as $\ket{\psi}_1=\hat{P}_{e1}\ket{\psi}=\ket{e_1}\ket{g_2}$.
Therefore, the determination of quantum state of qubit 1 being in the excited state immediately `pushes' the second qubit into the ground state. It is the central principle behind the quantum communication. Probability of getting the excited state for qubit 1 is $(1/N_2)^2$, as is the probability of getting the second qubit into the ground state. Likewise, the probability of obtaining the ground state of the first qubit is $1-(1/N_2)^2)$, which implies the same probability for the second qubit to be in the excited state.

Now, we are going to exercise the measurement of electron's position in the position-based qubit 1 by using the operator
$\hat{P}_{x_1} =\hat{I}_{cavity}\times (\ket{x_1}\bra{x_1}) \times \hat{I}_{qubit2}.$
 We assume that both qubits have the effective potential of the form $E_{p1}=E_{p2}=E_{p1'}=E_{p2'}=E_p$ and that $t_{s1}=t_{s2}$. In such a case, the individual energies of separated insulated qubits (1 or 2) are $E_{q1(q2)}=E_p \pm |t|$.
 Thus, we can write
$\hat{P}_{x_1} =\hat{I}_{cavity}\times \frac{1}{2}(\ket{g_1}+\ket{e_1})(\bra{g_1}+\bra{e_1}) \times \hat{I}_{qubit2}.$
Let us act this operator on qubit 1. Measurement of position of electron 1 is always due to an interaction of an external quantum system on qubit 1, so it is not surprising that this can change the energy.
We act with $\hat{P}_{x_1}$ on the quantum state $\ket{\psi}=\ket{E}_2$. We obtain
\begin{eqnarray}
\ket{\psi}_2=\nonumber \\=\hat{P}_{x_1}\ket{E_2}=(\hat{I}_{cavity}\times \frac{1}{2}(\ket{g_1}+\ket{e_1})(\bra{g_1}+\bra{e_1}) \times \hat{I}_{qubit2})(\frac{1}{N_2}(-(\frac{g_1}{g_2}e^{i(-d_2+d_1)})\ket{E_{\phi_1}}\ket{g_1}\ket{e_2}+\ket{E_{\phi_1}}\ket{e_1}\ket{g_2}))= \nonumber \\
=\frac{\ket{E_{\psi1}}}{2 N_2}(\ket{g_1}+\ket{e_1})(\bra{g_1}+\bra{e_1})(a \ket{g_1}\ket{e_2}+\ket{e_1}\ket{g_2}=\nonumber \\
\frac{\ket{E_{\psi1}}}{2 N_2}(-(\frac{g_1}{g_2}e^{i(-d_2+d_1)})(\ket{g_1}\ket{e_2}+\ket{e_1}\ket{g_2})+\ket{g_1}\ket{g_2}+\ket{e_1}\ket{g_2})=\nonumber \\
=\frac{1}{N_3}(-(\frac{g_1}{g_2}e^{i(-d_2+d_1)})(\ket{E_{\psi1}}\ket{g_1}\ket{e_2}+\ket{E_{\psi1}}\ket{e_1}\ket{g_2})+\ket{E_{\psi1}}\ket{g_1}\ket{g_2}+\ket{E_{\psi1}}\ket{e_1}\ket{g_2}).\nonumber \\
\end{eqnarray}
In a very real sense, the localization of electron in the position space brings its delocalization in the energy space. The quantum state after the measurement has a much richer energy spectrum than before the measurement. This is another demonstration of the uncertainty principle that is simply expressed by the fact that the more we know about the particle's position the less we know about its momentum, and reversely.

\section{Rigorous description of position-based qubit interacting with quantum cavity dynamics over time}

We are dealing with the quantum cavity interacting with the position-based qubit system that spans Hilbert space as
$\ket{\psi}_{qEC}\times \ket{\psi}_{qubit}$. At first, we are operating in the energy basis and so we can write
\begin{eqnarray}
\hat{H}=(\hat{H}_{qEC}\times \hat{I}_{qubit}+\hat{I}_{qEC} \times \hat{H}_{qubit})_0+\hat{E}_f \cdot \vec{p}_q+\hat{I}_{EC} \times \hat{V}(t)_{eff(qED \rightarrow qubit)}= \nonumber \\
\hat{I}_{EC} \times \hat{V}(t)_{eff(qED \rightarrow qubit)}+((E_{\phi1}\ket{E_{\phi1}}\bra{E_{\phi1}}+E_{\phi2}\ket{E_{\phi2}}\bra{E_{\phi2}})\times \hat{I}_{qubit}+\hat{I}_{qEC} \times (E_{g}\ket{E_{g}}\bra{E_{g}}+E_{e}\ket{E_{e}}\bra{E_{e}}))_0+ \nonumber \\.
+[(\ket{E_{\phi1}}\bra{E_{\phi1}}+\ket{E_{\phi2}}\bra{E_{\phi2}})\times (\ket{E_{g}}\bra{E_{g}}+\ket{E_{e}}\bra{E_{e}}) ] \hat{E}_f \cdot \vec{p}_q[(\ket{E_{\phi1}}\bra{E_{\phi1}}+\ket{E_{\phi2}}\bra{E_{\phi2}})\times (\ket{E_{g}}\bra{E_{g}}+\ket{E_{e}}\bra{E_{e}})] \nonumber \\
\end{eqnarray}
Let us consider the last term. It describes the interaction between the quantum electromagnetic cavity and the position-based qubit.
We have
\begin{eqnarray}
[(\ket{E_{\phi1}}\bra{E_{\phi1}}+\ket{E_{\phi2}}\bra{E_{\phi2}})\times (\ket{E_{g}}\bra{E_{g}}+\ket{E_{e}}\bra{E_{e}}) ] \hat{E}_f \cdot \vec{p}_q[(\ket{E_{\phi1}}\bra{E_{\phi1}}+\ket{E_{\phi2}}\bra{E_{\phi2}})\times (\ket{E_{g}}\bra{E_{g}}+\ket{E_{e}}\bra{E_{e}})]=\nonumber \\
=\ket{E_{\phi1}, E_{e}}(\bra{E_{\phi1},E_{e}} \hat{E}_f \cdot \vec{p}_q \ket{E_{\phi2}, E_{g}}) \bra{E_{\phi2}, E_{g}}+ \ket{E_{\phi2},E_{g}}(\bra{E_{\phi2},E_{g}} \hat{E}_f \cdot \vec{p}_q \ket{E_{\phi1},E_{e}}) \bra{E_{\phi1},E_{e}}.\nonumber \\
\end{eqnarray}
Here, we have explicitly neglected the terms $(\bra{E_{\phi1(2)},E_{g(e)}} \hat{E}_f \cdot \vec{p}_q \ket{E_{\phi1(2)}, E_{g(e)}})$ by setting them to zero, since they bring no energy exchange between the EM cavity and position based qubit, hence in that sense they are not physical.
Let us compute the term $(\bra{E_{\phi1},E_{e}} \hat{E}_f \cdot \vec{p}_q \ket{E_{\phi2}, E_{g}})$. We assume that the resulting Hamiltonian of interaction between the position based qubits and EM cavity is weak and can be treated as a perturbation.
We have 1- (3-)dimensional representation of EM field in the EM cavity (parameterized by $L$ or by ($L_1, L_2,L_3$)) given as
\begin{eqnarray}
\ket{E_{\phi1}}_t=\sqrt{\frac{\pi}{2L}}e^{\frac{1}{i \hbar}E_{\phi1}t}\int_{-\frac{L}{2}}^{+\frac{L}{2}}dx'\cos(\frac{x' \Pi}{L})\ket{x'}, \ket{E_{\phi2}}_t=\sqrt{\frac{2}{L}}e^{\frac{1}{i \hbar}E_{\phi2}t}e^{\frac{\hbar}{i}E_{\phi2}t}\int_{-\frac{L}{2}}^{+\frac{L}{2}}dx'\sin(\frac{ 2x' \Pi}{L})\ket{x'},
\end{eqnarray}
\begin{eqnarray}
\ket{E_{\phi1}}_{t,3}=(\sqrt{\frac{\pi}{2L_1}}e^{\frac{1}{i \hbar}E_{\phi1_a},t}\int_{-\frac{L_1}{2}}^{+\frac{L_1}{2}}dx'\cos(\frac{x' \Pi}{L_1})\ket{x'})(\sqrt{\frac{\pi}{2L_2}}e^{\frac{1}{i \hbar}E_{\phi1_b},t}\int_{-\frac{L_2}{2}}^{+\frac{L_2}{2}}dy'\cos(\frac{y' \Pi}{L_2})\ket{y'})\times \nonumber \\ \times (\sqrt{\frac{\pi}{2L_3}}e^{\frac{1}{i \hbar}E_{\phi1_c},t}\int_{-\frac{L_3}{2}}^{+\frac{L_3}{2}}dz'\cos(\frac{z' \Pi}{L_3})\ket{z'}), 
\end{eqnarray}
\begin{eqnarray}
\ket{E_{\phi2}}_{t,3}=(\sqrt{\frac{2}{L_1}}e^{\frac{1}{i \hbar}E_{\phi2_a},t}\int_{-\frac{L_1}{2}}^{+\frac{L_1}{2}}dx'\sin(\frac{2x' \Pi}{L_1})\ket{x'})(\sqrt{\frac{2}{L_2}}e^{\frac{1}{i \hbar}E_{\phi2_b},t}\int_{-\frac{L_2}{2}}^{+\frac{L_2}{2}}dy'\sin(\frac{2y' \Pi}{L_2})\ket{y'})\times \nonumber \\ \times (\sqrt{\frac{2}{L_3}}e^{\frac{1}{i \hbar}E_{\phi2_c},t}\int_{-\frac{L_3}{2}}^{+\frac{L_3}{2}}dz'\sin(\frac{z' 2\Pi}{L_3})\ket{z'}),
\end{eqnarray}
where $\ket{E_{\phi1(2)}}_t$ refers to the 1-dimensional case and $\ket{E_{\phi1(3)}}_{t,3}$ refers to the 3-dimensional case of EM cavity. In a very real way, the mathematical expression
\begin{eqnarray}
\psi_{E_{\phi1}}(x)_{t}=\bra{x}\ket{E_{\phi1}}_t=e^{\frac{E_{\phi_1}(t-t_0)}{\hbar i}}\psi_{E_{\phi1}}(x)_{t_0}=e^{\frac{E_{\phi_1}(t-t_0)}{\hbar i}}e^{i\alpha}\sqrt{\frac{\pi}{2L}} \cos(\frac{x \Pi}{L}), \nonumber \\
\psi_{E_{\phi2}}(x)_{t}=\bra{x}\ket{E_{\phi2}}_t=e^{\frac{E_{\phi_2}(t-t_0)}{\hbar i}}\psi_{E_{\phi2}}(x)_{t_0}=e^{\frac{E_{\phi_2}(t-t_0)}{\hbar i}}e^{i\beta}\sqrt{\frac{2}{L}} \sin(\frac{2 x \Pi}{L}), \nonumber \\
\psi_{E_{\phi1}}(x,y,z)_{t,3}=\bra{x}\bra{y}\bra{z}\ket{E_{\phi1}}_{t,3}=e^{\frac{E_{\phi_1}(t-t_0)}{\hbar i}}\psi_{E_{\phi1}}(x,y,z)_{t_0,3}=\nonumber \\
=e^{\frac{(E_{\phi1_a}+E_{\phi1_b}+E_{\phi1_c})(t-t_0)}{\hbar i}}e^{i\alpha}\sqrt{\frac{\pi}{2L_1}}\sqrt{\frac{\pi}{2L_2}}\sqrt{\frac{\pi}{2L_3}} \cos(\frac{x\Pi}{L_1})\cos(\frac{y\Pi}{L_2})\cos(\frac{z\Pi}{L_3}), \nonumber \\
\psi_{E_{\phi2}}(x,y,z)_{t,3}=\bra{x}\bra{y}\bra{z}\ket{E_{\phi2}}_{t,3}=e^{\frac{(E_{\phi_2a}+E_{\phi_2b}+E_{\phi_2c})(t-t_0)}{\hbar i}}\psi_{E_{\phi2}}(x,y,z)_{t_0,3}=\nonumber \\=e^{\frac{(E_{\phi2a}+E_{\phi2b}+E_{\phi2c})(t-t_0)}{\hbar i}}e^{i\beta}\sqrt{\frac{2}{L_1}}\sqrt{\frac{2}{L_2}}\sqrt{\frac{2}{L_3}} \sin(\frac{2 x \Pi}{L_1})\sin(\frac{2 y \Pi}{L_2})\sin(\frac{2 z \Pi}{L_3}), 
\end{eqnarray}
can be recognized as wavefunction $\psi_{E_{\phi1(2)}}(x)_{t} (\psi_{E_{\phi1(2)}}(x,y,z)_{t,3})$ of photon at mode 1(2) of electromagnetic cavity in 1- (3)-D case of size $L$ (or $L_1,L_2,L_3$ in 3-D rectangular cavity), where $\alpha,\beta \in R$.
In the case of 3-D cavity, the ground state wavefunction can be parameterized with wavefunction having three cosine functions, while the next excited state will have two cosine functions and one sinusoidal function.
The maximum concentration of energy (both electric and magnetic) is in the middle of cavity in the same fashion as with the quantum particle in a box. There are obvious analogies with the particle in a box and Schr\"odinger equation. However Schr\"odinger equation cannot be used for describing a photon confined in the EM cavity (since photon's rest mass is zero and it moves at the speed of light, which is not achievable for particle with non-zero rest mass and also momentum is proportional to energy in the case of photon) as resonating mode. Nevertheless, wavefunction of a photon in EC can be postulated. It is however possible to use the concept of Dirac (that is relativistically invariant under the Lorentz transformation and when we deal with wavefunction differential equation of the 1st order with respect to time and position derivative) equation for description of photon in EC and certain considerations in that direction are done by Bialynicki-Birula \cite{Birula}.
We limit our consideration to 1-D EC and we recognize that the quantum state of isolated, non-interacting with external world EM cavity, with two modes can be written as
\begin{equation}
\ket{E}_{cavity,t}=c_{E_{\phi1}}\ket{E_{\phi1}}_t+c_{E_{\phi2}}\ket{E_{\phi2}}_t, |c_{E_{\phi1}}|^2+|c_{E_{\phi2}}|^2=1(=n_1+n_2),
\end{equation}
where $|c_{E_{\phi1(2)}}|^2/(|c_{E_{\phi1}}|^2+|c_{E_{\phi2}}|^2)$ are the probabilities of occupancy of 1(2) resonating modes and $n_{1(2)}$ is the number of photos populating the 1st or 2nd mode. We recognize than a minimal number of $n_1+n_2$ is one photon. Minimum energy necessary to fully populate mode 1(2) of EC is $E_{cavity1(2)}=\hbar \frac{2(4)\Pi}{L}$. By having qEC described in the energy basis, it quite easy to conduct a measurement of its energy. If the outcome of measurement indicates that the cavity has energy $E_{\phi1(2)}$, then the act of measurement can be represented by the operator $\hat{P}_{E\phi1(2)}$=$\ket{E_{\phi1(2)}}\bra{E_{\phi1(2)}}$
acting on the quantum state $\ket{E}_{cavity,t}$ at time instant $t$
so the resulting state is $\ket{E}_{cavityQ,t}=\frac{\hat{P}_{E_{\phi1(2)}}\ket{E}_{cavity,t}}{\bra{E}_{cavity,t} \hat{P}_{E\phi1(2)} \ket{E}_{cavity,t}}$. 
Much more involving is the measurement conducted on the quantum cavity when one determines the presence of photon in the space interval $x \in (a_s,b_s) \in (-L/2, L/2)$, as can be obtained by some detector of EM field. In such a case, the measurement is represented by the measurement projector
$\hat{P}(a_s,b_s)=\int_{a_s}^{b_s}dx'''\ket{x'''}\bra{x'''}$. Again, we have a collapse of quantum state after the measurement at time instant $t$, so the quantum state becomes $\ket{E}_{cavityQ',t}=\frac{\int_{a_s}^{b_s}dx'''\ket{x'''}\bra{x'''}\ket{E}_{cavity,t}}{\bra{E}_{cavity,t} \int_{a_s}^{b_s}dx'''\ket{x'''}\bra{x'''} \ket{E}_{cavity,t}}$. We notice that given the representation of position measurement allows us to conduct the measurement of photon position at $x_p$, when $a_s \rightarrow b_s=x_p$. Alternatively, a projector operator corresponding to the measurement of photon at $x=x_p$ can be written as $\hat{P}(x_p)=\int_{-\infty}^{-\infty}dx'''\ket{x'''}\bra{x'''}\delta(x'''-x_p)$.
In similar fashion we can exercise weak measurements both in the energy or position bases, but it requires the linear combination of projectors (either in the energy or position basis).

It is quite straightforward to generalize all the conducted reasoning for $m$ modes of EM cavity, but for the sake of simplicity we consider two modes of resonating cavity.
It is worth underlining that the operator $a^{\dag}=\ket{E_{\phi2}}\bra{E_{\phi1}}$ describes the act of populating state $\ket{E_{\phi_2}}$ and depopulating state $\ket{E_{\phi_1}}$ so it describes the act of delivering the energy to the EM cavity, while
$\hat{a}=\ket{E_{\phi1}}\bra{E_{\phi2}}$ describes the act of populating the state of lower energy $\ket{E_{\phi_1}}$ and depopulation the state of higher energy $\ket{E_{\phi_2}}$, so it describes the act of the EM energy decrease by the electromagnetic cavity. In such a case, we can describe the dissipation processes in language of quantum mechanics so in general case non-Hermitian Hamiltonian that is able to deliver or remove the energy from the EM cavity is of the form $\hat{H}_{disspation EC}=f_1(t)\ket{E_{\phi2}}\bra{E_{\phi1}}+f_2(t)\ket{E_{\phi1}}\bra{E_{\phi2}}$, where $f_1(t), f_2(t)$ are complex-valued functions. If $f_1(t)=f_2(t)^{*}$, then we deal with a Hermitian Hamiltonian and our EM cavity is subjected to Rabi oscillations between energy levels $E_{\phi1}$ and $E_{\phi2}$.

We are using an analogy of the quantum EM cavity with a quantum harmonic oscillator. In qEC, the electric field plays a role of particle position in the q-harmonic oscillator and the magnetic field is analogous to the momentum of particle in the harmonic oscillator described by the Schr\"odinger equation. Those analogies give a hint that both the electric and magnetic field operators in qEC shall be linear combinations of creation and annihilation operators denoted as $\ket{E_{\phi_2}}\bra{E_{\phi_1}}$ and $\ket{E_{\phi_1}}\bra{E_{\phi_2}}$. Indeed, in the case of simplistic quantum harmonic oscillator approach, we have
$\hat{a}=\frac{1}{\sqrt{2}}(\hat{x}+i\hat{p}), \hat{a}^{\dag}=\frac{1}{\sqrt{2}}(\hat{x}-i\hat{p})$, so $\frac{1}{\sqrt{2}}(\hat{a}+\hat{a}^{\dag})=\hat{x}$ and $\frac{i}{\sqrt{2}}(\hat{a}^{\dag}-\hat{a})=\hat{p}$. In the case of harmonic oscillator, we have $\hat{H}=\hbar \omega(\hat{a}^{\dag}\hat{a}+\frac{1}{2})=\frac{1}{2}(\hat{x}^{2}+\hat{p}^{2})$. In the same fashion, we expect that the electric field and magnetic field operators will give $\hat{H}_{qEC}=\frac{1}{2}(\hat{E}_f^{\dag}\hat{E}_f+\hat{B}_f^{\dag}\hat{B}_f)$.
Let us state a hypothesis: $\frac{1}{\sqrt{2}}r_1(x)o_1(t)(\hat{a}^{\dag}+\hat{a})=\frac{1}{\sqrt{2}}r_1(x)o_1(t)(\ket{E_2}\bra{E_1}+\ket{E_1}\bra{E_2})=\hat{E}(x,t)_f$ and $\frac{i}{\sqrt{2}}r_2(x)o_2(t)(\hat{a}^{\dag}-\hat{a})=\frac{i}{\sqrt{2}}r_2(x)o_2(t)(\ket{E_2}\bra{E_1}-\ket{E_1}\bra{E_2})=\hat{B}(x,t)_f$. 

Now, we examine the structure of operator $\hat{E}_f^{\dag}\hat{E}_f=\frac{1}{2}(r_1(x))^2(o_1(t))^2(\ket{E_2}\bra{E_2}+\ket{E_1}\bra{E_1})=\frac{1}{2}(r_1(x))^2(o_1(t))^2\hat{1}$. After this, we check the structure of operator $\hat{B}_f^{\dag}\hat{B}_f=\frac{1}{2}(r_2(x))^2(o_2(t))^2(\ket{E_1}\bra{E_2}-\ket{E_2}\bra{E_1})(\ket{E_2}\bra{E_1}-\ket{E_1}\bra{E_2})=\frac{1}{2}(r_2(x))^2(o_2(t))^2\hat{1}$. It shall be pointed out that the total energy of electrostatic and magnetic fields at any point of EC is not changing over time, which suggests $\frac{1}{2}(r_2(x))^2(o_2(t))^2+\frac{1}{2}(r_1(x))^2(o_1(t))^2=u(x)$. One of the solutions is that $constant=u=(o_2(t))^2+(o_1(t))^2$ and this suggests $o_1(t)=u_0 sin(\omega_1t+p_1)$ and $o_2(t)=u_0 cos(\omega_1+p_1)$, and at the same time $r_1(x)=r_2(x)=E_0 cos(\frac{2\pi}{L}x)$.
We postulate that for $k$-th mode of qEC, the electric field operator shall be $\hat{E}(k)_f=(\ket{E_{k+1}}\bra{E_{1}}+\ket{E_{1}}\bra{E_{k+1}})g(x)_k E_{ox,k} \sqrt{E_0}$ and $\hat{B}(k)_f=i(\ket{E_{k+1}}\bra{E_{1}}-\ket{E_{1}}\bra{E_{k+1}})\sqrt{1-|g(x)_k|^2} \sqrt{E_0}$ and obviously the electric field operator corresponds to the perpendicular magnetic field, and the photon momentum is perpendicular to both the electric and magnetic fields as it is proportional to the Poyting vector, hence it is proportional to $\vec{E}\times \vec{B}$.
We can introduce the generalized electric field operator having m modes of qEC expressed in the form:
$\hat{E_f}(t)=\sum_{k=2}^{k=m}\hat{E_f}(t)=(\ket{E_{k+1}}\bra{E_{1}}+\ket{E_{1}}\bra{E_{k+1}})E_{ox,k}g(x)_k$.
Now we introduce the electric field operator of quantum electromagnetic cavity in the space representation as
\begin{eqnarray}
(\int \ket{x''}\bra{x''}dx'')\hat{E}(x_s,t)_f(\int \ket{x'}\bra{x'}dx')=\nonumber \\ =\int_{-\infty}^{+\infty}\int_{-\infty}^{+\infty}\int_{-\infty}^{+\infty}dxdx' dx''cos(\frac{x \Pi}{L})sin(\frac{E_{\phi1}}{\hbar}t+\gamma_1)
[\sqrt{\frac{2}{L}}\sqrt{\frac{\pi}{2L}}[E_{ox,1}'cos(\frac{x'' \Pi}{L})sin(\frac{ 2x' \Pi}{L})+ \nonumber \\
 +E_{ox,2}'sin(\frac{ 2x'' \Pi}{L})cos(\frac{x' \Pi}{L}] \delta(x-x_s)\ket{x''}\bra{x'}].
\end{eqnarray}
The conducted reasoning is given for the 1-D case but quite easily it can be extended to the 2- and 3-D description of EM cavity.
Maximum strength of electric field in electromagnetic cavity for modes 1 and 2 is encoded by coefficients $E_{ox,1}$ and $E_{ox,2}$. It shall be underlined that $|E_{ox,1(2)}|^2$ is proportional to the number of photons $n_1(1)$ in mode 1(2). Since photons are
bosons, we can have an arbitrarily large number of photons at any mode 1 or 2, or $n$-th. In a very real sense, it is a manifestation of Bose-Einstein condensation of photons.

For the case of position-based qubit, we can define the electric diplole moment of electrostatic qubit to be of the form $\hat{P}_f=\frac{\vec{d}}{2}(\ket{x_1}\bra{x_1}-\ket{x_2}\bra{x_2}) e$, as the difference between electric charge present at nodes 1 and 2 contributes to the dipole moment, where $d$ is the distance between points 1 and 2 in the qubit and $e$ is the electron charge. We can assume the qubit point-like structure so we have the effective electric field as an average between points 1 and 2, which can be formally written as
$\hat{E}_f(x_{qubit},t)=\frac{\hat{E}(\frac{x_1+x_2}{2},t)_f}{2}(\ket{x_1}\bra{x_1}+\ket{x_2}\bra{x_2})$ , where $x_{qubit}=\frac{x_1+x_2}{2}$. We define $a,b,c,d$ coefficients as $\ket{E}_g=\ket{E}_g=a\ket{x_1}+b\ket{x_2}$ and $\ket{E}_g=\ket{E}_g=a\ket{x_1}+b\ket{x_2}$. This brings the following structure of the operator $\hat{E}_f(x_{qubit},t)\hat{P}_f=\frac{\hat{E}_f(x_1,t)_f+\hat{E}_f(x_2,t)_f}{2}\frac{d}{2}(\ket{x_1}\bra{x_1}-\ket{x_2}\bra{x_2})$=$(\ket{E_{\phi2}}_t\bra{E_{\phi1}}_t+\ket{E_{\phi1}}_t\bra{E_{\phi2}}_t)e E_{ox,k}\frac{(g(x_1)_k+g(x_2)_k)}{2}\frac{x_2-x_1}{2}(\ket{E_g}\bra{E_g}+\ket{E_e}\bra{E_e})(\ket{x_1}\bra{x_1}-\ket{x_2}\bra{x_2})(\ket{E_g}\bra{E_g}+\ket{E_e}\bra{E_e}) $. We obtain $(\ket{E_g}\bra{E_g}+\ket{E_e}\bra{E_e})(\ket{x_1}\bra{x_1}-\ket{x_2}\bra{x_2})(\ket{E_g}\bra{E_g}+\ket{E_e}\bra{E_e})=\ket{E_g}\bra{E_g}(|a|^2-|b|^2)+\ket{E_e}\bra{E_e}(|c|^2-|d|^2)$+$\ket{E_g}\bra{E_e}(a^{*}c-b^{*}d)+\ket{E_e}\bra{E_g}(ac^{*}-bd^{*})$.

We can write $(\bra{E_{\phi1},E_{e}} \hat{E}_f \cdot \vec{p}_q \ket{E_{\phi2}, E_{g}})$ and $(\bra{E_{\phi2},E_{g}} \hat{E}_f \cdot \vec{p}_q \ket{E_{\phi1}, E_{e}})$ and the interaction qubit-qEC Hamiltonian is given as
\begin{eqnarray}
\label{CenterQ}
(\bra{E_{\phi1},E_{e}} \hat{E}_f \cdot \vec{p}_q \ket{E_{\phi2}, E_{g}})_{t}=(\bra{E_{e}} \vec{p}_q \ket{E_{g}})_{t}(\bra{E_{\phi1}} \hat{E}_f \ket{E_{\phi2}})_{t}=\nonumber \\ =(\bra{E_{e,t0}} \vec{p}(t)_q \ket{E_{g,t0}})\ket{E_{\phi2}})_{t0}(\bra{E_{\phi1,t0}} \hat{E}_{f,t0} \ket{E_{\phi2,t0}}) 
e^{\frac{1}{i \hbar}(E_{\phi2}-E_{\phi1})(t-t_0)}e^{\frac{1}{i \hbar}(E_{g}-E_{e})(t-t_0)} =\nonumber \\
=(\bra{E_{e}} \vec{p}_q \ket{E_{g}})_t 
e^{\frac{1}{i \hbar}(E_{\phi2}-E_{\phi1})(t-t_0)}E_{ox,k}\frac{(g(x_1)_k+g(x_2)_k)}{2}=\nonumber \\=(ac^{*}-bd^{*})_{t}\frac{e(x_2-x_1)}{4}(E_{ox,k}\frac{(g(x_1)_k+g(x_2)_k)}{2})_{t_0}e^{\frac{1}{i \hbar}(E_{\phi2}-E_{\phi1})(t-t_0)}e^{\frac{1}{i \hbar}(E_{g}-E_{e})(t-t_0)}=u(t),\nonumber \\
(\bra{E_{\phi1},E_{g}} \hat{E}_f \cdot \vec{p}_q \ket{E_{\phi1}, E_{e}})_{t}=u^{*}(t), \nonumber \\
\hat{H}_{qEC-qubit}=\ket{E_{\phi1},E_{e}}(\bra{E_{\phi1},E_{e}} \hat{E}_f \cdot \vec{p}_q \ket{E_{\phi2}, E_{g}})_{t}\bra{E_{\phi2},E_{g}}+\ket{E_{\phi2},E_{g}}(\bra{E_{\phi2},E_{g}} \hat{E}_f \cdot \vec{p}_q \ket{E_{\phi1}, E_{e}})_{t}\bra{E_{\phi1}, E_{e}})+\nonumber \\+
(\ket{E_{\phi1},E_{g}}\bra{E_{\phi2},E_{g}}+\ket{E_{\phi2},E_{g}}\bra{E_{\phi1},E_{g}})_{t}\frac{|a_{t}|^2-|b_{t}|^2}{2}E_{ox,k}\frac{(g(x_1)_k+g(x_2)_k)}{2}ed+\nonumber \\+(\ket{E_{\phi1},E_{e}}\bra{E_{\phi2},E_{e}}+\ket{E_{\phi2},E_{e}}\bra{E_{\phi1},E_{e}})_{t}\frac{|c_{t}|^2-|d_{t}|^2}{2}E_{ox,k}\frac{(g(x_1)_k+g(x_2)_k)}{2}ed\nonumber \\
=\ket{E_{\phi1},E_{e}} u(t) \bra{E_{\phi2},E_{g}}+\ket{E_{\phi2},E_{g}} u(t)^{*} \bra{E_{\phi1},E_{e}}\nonumber
+b_1(t)(\ket{E_{\phi1},E_{g}}\bra{E_{\phi2},E_{g}}+\ket{E_{\phi2},E_{g}}\bra{E_{\phi1},E_{g}})_{t}+\nonumber\\
+b_2(t)(\ket{E_{\phi1},E_{e}}\bra{E_{\phi2},E_{e}}+\ket{E_{\phi2},E_{e}}\bra{E_{\phi1},E_{e}})_{t}, \nonumber \\
b_1(t)=ed\frac{|a_{t}|^2-|b_{t}|^2}{2}E_{ox,k}\frac{(g(x_1)_k+g(x_2)_k)}{2}, b_2(t)=ed\frac{|c_{t}|^2-|d_{t}|^2}{2}E_{ox,k}\frac{(g(x_1)_k+g(x_2)_k)}{2}.\nonumber \\
\end{eqnarray}

It is convenient to write the matrix corresponding to the global qEC-qubit Hamiltonian 
in the form
\begin{eqnarray}
\hat{H}=
\begin{pmatrix}
E_{\phi1}+E_{g}(t) & 0 & b_1(t) & 0 \\
0 & E_{\phi1}+E_{e}(t) & (\bra{E_{\phi1},E_{e}} \hat{E}_f \cdot \vec{p}_q \ket{E_{\phi2}, E_{g}})_{t} & b_2(t) \\
b_1(t)^{*} & (\bra{E_{\phi2},E_{g}} \hat{E}_f \cdot \vec{p}_q \ket{E_{\phi1}, E_{e}})_{t} & E_{\phi2}+E_{g}(t) & 0 \\
0 & b_2(t)^{*} & 0 & E_{\phi2}+E_{e}(t) \\
\end{pmatrix}.
\end{eqnarray}

Since the world related to the technology is closer to the description in the position space (assuming $t_{s12(t)}=|t_s(t)|e^{i \alpha(t)}$), we have
\begin{eqnarray}
\ket{E_g,t}_n=
\frac{((E_{p2}(t)-E_{p1}(t))+\sqrt{\frac{(E_{p2}(t)-E_{p1}(t))^2}{2}+|t_{s12}(t)|^2})e^{i \alpha(t)}i\ket{x_1}-|t_s(t)|\ket{x_2})}{\sqrt{|t_s(t)|^2+((E_{p2}(t)-E_{p1}(t))+\sqrt{\frac{(E_{p2}(t)-E_{p1}(t))^2}{2}+|t_{s12}(t)|^2})^2}}
=a(t)\ket{x_1}_n+b(t)\ket{x_2}_n,
\nonumber \\
\ket{E_e,t}_n=
\frac{(-(E_{p2}(t)-E_{p1}(t))+\sqrt{\frac{(E_{p2}(t)-E_{p1}(t))^2}{2}+|t_{s}(t)|^2})e^{-i\alpha(t)}\ket{x_1}+|t_s(t)|\ket{x_2})}{\sqrt{|t_s|^2+(-(E_{p2}(t)-E_{p1}(t))+\sqrt{\frac{(E_{p2}(t)-E_{p1}(t))^2}{2}+|t_{s12}(t)|^2})^2}}=c(t)\ket{x_1}_n+d(t)\ket{x_2}_n, 
\nonumber \\
a(t)=\frac{((E_{p2}(t)-E_{p1}(t))+\sqrt{\frac{(E_{p2}(t)-E_{p1}(t))^2}{2}+|t_{s12}(t)|^2})e^{i \alpha(t)}i}{\sqrt{|t_s(t)|^2+((E_{p2}(t)-E_{p1}(t))+\sqrt{\frac{(E_{p2}(t)-E_{p1}(t))^2}{2}+|t_{s12}(t)|^2})^2}}, \nonumber \\
b(t)=\frac{-|t_s(t)|}{\sqrt{|t_s(t)|^2+((E_{p2}(t)-E_{p1}(t))+\sqrt{\frac{(E_{p2}(t)-E_{p1}(t))^2}{2}+|t_{s12}(t)|^2})^2}}, \nonumber \\
c(t)=\frac{-(E_{p2}(t)-E_{p1}(t))+\sqrt{\frac{(E_{p2}(t)-E_{p1}(t))^2}{2}+|t_{s}(t)|^2})e^{-i\alpha(t)}}{\sqrt{|t_s|^2+(-(E_{p2}(t)-E_{p1}(t))+\sqrt{\frac{(E_{p2}(t)-E_{p1}(t))^2}{2}+|t_{s12}(t)|^2})^2}}, \nonumber \\
d(t)=\frac{+|t_s(t)|}{\sqrt{|t_s|^2+(-(E_{p2}(t)-E_{p1}(t))+\sqrt{\frac{(E_{p2}(t)-E_{p1}(t))^2}{2}+|t_{s12}(t)|^2})^2}},
\nonumber \\
|a(t)|^2+|b(t)|^2=1, |c(t)|^2+|d(t)|^2=1, b(t) \in R, d(t) \in R.
\end{eqnarray}
Now, we investigate the algebraic structure of operators $\ket{E_1,t}_n\bra{E_2,t}_n$ and $\ket{E_2,t}_n\bra{E_1,t}_n$ in terms of operators $\ket{x_1}_n\bra{x_2}_n$, $\ket{x_2}_n\bra{x_1}_n$, $\ket{x_1}_n\bra{x_1}_n$ , $\ket{x_2}_n\bra{x_2}_n$. We can write the eigenenergy operators of qubit in terms of the position representation as
   \begin{eqnarray}
\ket{E_1,t}_n\bra{E_2,t}_n=(a(t)\ket{x_1}_n+b(t)\ket{x_2}_n)(c^{*}(t)\bra{x_1}_n+d^{*}(t)\bra{x_2}_n)= \nonumber \\
=a(t)c^{*}(t)\ket{x_1}_n\bra{x_1}_n+ b(t)d^{*}(t)\ket{x_2}_n\bra{x_2}_n+a(t)d^{*}(t)\ket{x_1}_n\bra{x_2}_n+b(t)c^{*}(t)\ket{x_2}_n\bra{x_1}_n
   \end{eqnarray}
   \begin{eqnarray}
\ket{E_2,t}_n\bra{E_1,t}_n=(c(t)\ket{x_1}_n+d(t)\ket{x_2}_n)(a^{*}(t)\bra{x_1}_n+b^{*}(t)\bra{x_2}_n)=\nonumber \\
=a(t)^{*}c(t)\ket{x_1}_n\bra{x_1}_n+ b(t)^{*}d(t)\ket{x_2}_n\bra{x_2}_n+a(t)^{*}d(t)\ket{x_1}_n\bra{x_2}_n+b(t)^{*}c(t)\ket{x_2}_n\bra{x_1}_n.
   \end{eqnarray}
Now let us investigate $(\bra{E_{e}} \vec{p}_q \ket{E_{g}})$ and $(\bra{E_{g}} \vec{p}_q \ket{E_{e}})$.
We obtain
\begin{eqnarray}
(\bra{E_{e}} \vec{p}_q \ket{E_{g}})=((c^{*}(t)\bra{x_1}_n+d^{*}(t)\bra{x_2}_n)(\frac{1}{2}(\ket{x_1}\bra{x_1}-\ket{x_2}\bra{x_2})d e)((a(t)\ket{x_1}_n+b(t)\ket{x_2}_n)=\nonumber \\
=\frac{1}{2}(c^{*}(t)a(t)-d^{*}(t)b(t))=u(t), \nonumber \\
(\bra{E_{g}} \vec{p}_q \ket{E_{e}})=((a^{*}(t)\bra{x_1}_n+b^{*}(t)\bra{x_2}_n)(\frac{1}{2}(\ket{x_1}\bra{x_1}-\ket{x_2}\bra{x_2})d e)((c(t)\ket{x_1}_n+d(t)\ket{x_2}_n)=\nonumber \\
=\frac{1}{2}(c(t)a(t)^{*}(t)-d(t)b^{*}(t))=u(t)^{*}.
\end{eqnarray}
which implies that qEC-qubit interaction Hamiltonian is of the form
\begin{eqnarray}
\hat{H}_{qEC-qubit}=\nonumber \\=\ket{E_{\phi1}}\ket{E_{e}}(\bra{E_{e}} \vec{p}_q \ket{E_{g}})(\bra{E_{\phi1}} \hat{E}_f \ket{E_{\phi2}})\bra{E_{\phi2}}\bra{E_{g}}+\ket{E_{\phi2}}\ket{E_{g}}(\bra{E_{g}} \vec{p}_q \ket{E_{e}})(\bra{E_{\phi2}} \hat{E}_f \ket{E_{\phi1}})\bra{E_{\phi1}}\bra{E_{e}}=\nonumber \\
=\frac{1}{2}\ket{E_{\phi1}}(c(t)\ket{x_{1}}+d(t)\bra{x_{2}})(c^{*}(t)a(t)-d^{*}(t)b(t))(\bra{E_{\phi1}} \hat{E}_f \ket{E_{\phi2}})\bra{E_{\phi2}}(a^{*}(t)\bra{x_{1}}+b^{*}(t)\bra{x_{2}})+\nonumber \\
+\frac{1}{2}\ket{E_{\phi2}}(a(t)\ket{x_{1}}+b(t)\ket{x_{2}})(c(t)a(t)^{*}-d(t)b^{*}(t))(\bra{E_{\phi2}} \hat{E}_f \ket{E_{\phi1}})\bra{E_{\phi1}}(c^{*}(t)\bra{x_{1}}+d^{*}(t)\bra{x_{2}})=\nonumber \\
=\frac{1}{2}\Bigg[\ket{E_{\phi2}}\ket{x_{1}}(\bra{E_{\phi2}}  \hat{E}_f \ket{E_{\phi1}})(|a(t)|^2|c(t)|^2-a(t)d(t)b^{*}(t)c^{*}(t))\bra{E_{\phi1}}\bra{x_{1}} +\nonumber \\
+\ket{E_{\phi1}}\ket{x_{1}}(\bra{E_{\phi1}}  \hat{E}_f \ket{E_{\phi2}})(|a(t)|^2|c(t)|^2-a^{*}(t)d^{*}(t)b(t)c(t))\bra{E_{\phi2}}\bra{x_{1}}+\nonumber \\
+\ket{E_{\phi2}}\ket{x_{2}}(\bra{E_{\phi2}}  \hat{E}_f \ket{E_{\phi1}})(-|b(t)|^2|d(t)|^2+b(t)c(t)a(t)^{*}d(t)^{*})\bra{E_{\phi1}}\bra{x_{2}} +\nonumber \\
+\ket{E_{\phi1}}\ket{x_{2}}(\bra{E_{\phi1}}  \hat{E}_f \ket{E_{\phi2}})(-|b(t)|^2|d(t)|^2+b(t)^{*}c(t)^{*}a(t)d(t))\bra{E_{\phi2}}\bra{x_{2}}+\nonumber \\
+\ket{E_{\phi2}}\ket{x_{2}}(\bra{E_{\phi2}}  \hat{E}_f \ket{E_{\phi1}})(|c(t)|^2b(t)a^{*}(t)-d(t)c^{*}(t)|b(t)|^2)\bra{E_{\phi1}}\bra{x_{1}}+\nonumber \\
+\ket{E_{\phi1}}\ket{x_{2}}(\bra{E_{\phi1}}  \hat{E}_f \ket{E_{\phi2}})(d(t)|a(t)|^2c^{*}(t)-|d(t)|^2b(t)a(t)^{*})\bra{E_{\phi2}}\bra{x_{1}}+\nonumber \\
+\ket{E_{\phi2}}\ket{x_{1}}(\bra{E_{\phi2}}  \hat{E}_f \ket{E_{\phi1}})(c(t)d(t)^{*}|a(t)|^2-|d(t)|^2 a(t)b^{*}(t))\bra{E_{\phi1}}\bra{x_{2}}+\nonumber \\
+\ket{E_{\phi1}}\ket{x_{1}}(\bra{E_{\phi1}}  \hat{E}_f \ket{E_{\phi2}})(a(t)b^{*}(t)|c(t)|^2-c(t)d(t)^{*}|b(t)|^2)\bra{E_{\phi2}}\bra{x_{2}}]\Bigg.\nonumber \\
\end{eqnarray}
The last operator $\hat{H}_{qEC-qubit,\ket{E_{\phi}}\ket{x_q}}$ has the following matrix representation in the basis $\ket{E_{\phi1}}\ket{x_1}$,$\ket{E_{\phi1}}\ket{x_2}$, $\ket{E_{\phi2}}\ket{x_1}$,$\ket{E_{\phi2}}\ket{x_2}$ as
\small
\begin{eqnarray}
\hat{H}_{qEC-qubit}=\frac{1}{2} 
\begin{pmatrix}
0 & 0 & u_t(|a_t|^2|c_t|^2-a^{*}_td^{*}_tb_tc_t) & u_t(a_tb^{*}_t|c_t|^2-c_td_t^{*}|b_t|^2) \\
0 & 0 & u_t(d_t|a_t|^2c^{*}_t-|d_t|^2b_ta_t^{*})& u_t(-|b_t|^2|d_t|^2+b_t^{*}c_t^{*}a_td_t) \\
u^{*}_{t}(|a_t|^2|c_t|^2-a_td_tb^{*}_tc^{*}_t) & u^{*}_t(c_td_t^{*}|a_t|^2-|d_t|^2 a_tb^{*}_t) & 0 & 0 \\
u^{*}_t(|c_t|^2b_ta^{*}_t-d_tc^{*}_t|b_t|^2) & u_t^{*}(-|b_t|^2|d_t|^2+b_tc_ta_t^{*}d_t^{*}) & 0 & 0 \\
\end{pmatrix}. 
\end{eqnarray}
\normalsize

\section{Renormalization of qubit and resonant cavity Hamiltonians due to mutual phase imprints and due to electrostatic modification of qubit effective potential}

From electrodynamics, it is known that moving a charge (e.g. electron) generates a non-zero vector potential and thus non-zero magnetic field that can build-up phase imprint on the quantum electromagnetic cavity. On the other hand, the oscillating electromagnetic cavity generates a non-zero vector potential (and thus magnetic field) that brings a phase imprint on the electron confined in the electrostatic position-based qubit or any other qubit type. The measure of electron kinetic energy and its momentum are encoded in the hopping parameter in the tight-binding model $t_s(time)$ which generates a vector potential proportional to $|t_s(time)|$. The measure of vector potential generated by electromagnetic cavity is described by a frequency of oscillating electric and magnetic fields, since from electrodynamics we have the relation $-\vec{\nabla} V(time,x,y,z)-\frac{d}{c dt}\vec{A}(time,x,y,z)_x=\vec{E}(time,x,y,z)$ and we set $V=0$ (a special choice of vector potential gauge), where $c$ is the speed of light. Also we encounter the situation when local confinement potential governing the electron confinement at two nodes is modified by an external electrostatic potential generated by a time dependent electric field coming from EC. Described effects are quite small, nevertheless they take place in the real physical situation. Let us examine the structure of Hamiltonian of isolated non-interacting EM cavity and isolated qubit. We have
\begin{eqnarray}
\hat{H}_{\ket{E_{\phi}}\ket{x}}=(\hat{H}_{qEC}\times \hat{I}_{qubit}+\hat{I}_{qEC} \times \hat{H}_{qubit})_0
=\begin{pmatrix}
E_{\phi1} & 0 & 0 & 0 \\
0 & E_{\phi1} & 0 & 0 \\
0 & 0 & E_{\phi2} & 0 \\
0 & 0 & 0 & E_{\phi2} \\
\end{pmatrix}+
\begin{pmatrix}
E_{p1}(t) & t_s(t) & 0 & 0 \\
t_s(t)^{*} & E_{p2}(t) & 0 & 0 \\
0 & 0 & E_{p1}(t) & t_s(t) \\
0 & 0 & t_s(t)^{*} & E_{p2}(t) \\
\end{pmatrix}
\end{eqnarray}
Equations of motion of considered closed physical system are $\hat{H(t)}\ket{\psi(t)}=i\hbar\frac{d}{dt}\ket{\psi(t)}$ and imply \newline $\ket{\psi(t)}=e^{\frac{\int_{t_0}^{t}\hat{H}(t')dt'}{i\hbar}}\ket{\psi(t_0)}=\hat{U}(t,t_0)\ket{\psi(t_0)}$.

 We are dealing with the following quantum state Hilbert space that is a mixture of energy-position representation of qEC-qubit physical system given as
\begin{equation}
\ket{\psi(t)}=c_1(t)\ket{E_{\phi1}}\ket{x_1}+c_2(t)\ket{E_{\phi1}}\ket{x_2}+c_3(t)\ket{E_{\phi2}}\ket{x_1}+c_4(t)\ket{E_{\phi2}}\ket{x_2}, |c_1(t)|^2+..+|c_4(t)|=1.
\end{equation}
Formal definition of the hopping term in the tight-binding model can be given as $\bra{x_1} \hat{H}_{qubit}(t)\ket{x_2}$=$t_{s21}(t)$, $t_{s12}(t)=t_{s21}^{*}(t)$, $\bra{x_2} \hat{H}_{qubit}(t)\ket{x_1}$=$t_{s12}(t)$. On the other hand, the formal definition of electrostatic energy in case of qED can be given as $\bra{E_{\phi1}} \hat{E}_{f}(x_{qubit},t)\ket{E_{\phi1}}$, $\bra{E_{\phi2}}\hat{E}_{f}(x_{qubit},t) \ket{E_{\phi2}}$. There is an anticorrelation in the mutual phase imprints that is on the qubit from the electromagnetic cavity and that on the electromagnetic cavity from the qubit. This mutual phase imprint is minimizing the electric and magnetic energy of coupling qubit-qEC system. It has its thermodynamics justification since any given physical system has always such equations of motion that tend to bring the system to an equilibrium and to a possible ground state or to a dynamic equilibrium as it is the case of time crystals considered by Wilczek \cite{Wilczek} and Sacha \cite{Sacha}. We can postulate the following structure of renormalized Hamiltonian matrix incorporating phase imprint generated on qubit from qEC.

\begin{eqnarray}
\hat{H}_{(\ket{E_{\phi}}\ket{x})_{r1,0}}=(\hat{H}_{qEC}\times \hat{I}_{qubit})_{r1,0}+(\hat{I}_{qEC} \times \hat{H}_{qubit})_{r1,0} 
=\begin{pmatrix}
E_{\phi1} & 0 & 0 & 0 \\
0 & E_{\phi1} & 0 & 0 \\
0 & 0 & E_{\phi2} & 0 \\
0 & 0 & 0 & E_{\phi2} \\
\end{pmatrix}_{r1,0}+\nonumber \\
\begin{pmatrix}
E_{p1}(t) & t_s(t)e^{is_0 \int_{0}^{t} dt'\bra{E_{\phi1}} \hat{E}_{f}(x_{q},t')\ket{_{\phi1}}} & 0 & 0 \\
t_s(t)^{*}e^{-is_0  \int_{0}^{t} dt'\bra{E_{\phi1}} \hat{E}_{f}(x_{q},t')\ket{E_{\phi1}}} & E_{p2}(t) & 0 & 0 \\
0 & 0 & E_{p1}(t) & t_s(t)e^{is_0 \int_{0}^{t} dt'\bra{E_{\phi2}} \hat{E}_{f}(x_{q},t')\ket{E_{\phi2}}} \\
0 & 0 & t_s(t)^{*}e^{-is_0 \int_{0}^{t} dt'\bra{E_{\phi2}} \hat{E}_{f}(x_{q},t')\ket{E_{\phi2}}} & E_{p2}(t) \\
\end{pmatrix}_{r1,0}
\end{eqnarray}
\normalsize
Taking into account the second renormalization coming from the fact that a moving electron is generating phase imprint on qEC can be only accounted if we consider the unitarian evolution operator.
Now it is time to take into account the modification of local-confinement qubit potential by oscillating electromagnetic cavity. We can assume modification of diagonal elmements as $V_{qED_{\phi_1} \rightarrow qubit}(1,t)=V_{a1}(E_{ox,1}')\sin(\frac{E_{\phi1}t}{\hbar}+p_1), V_{qED_{\phi_2} \rightarrow qubit}(1,t)= V_{a2}(E_{ox,2}')\sin(\frac{E_{\phi2}t}{\hbar}+p_2)$,
$V_{qED_{\phi_1} \rightarrow qubit}(2,t)=V_{a3}(E_{ox,1}')\sin(\frac{E_{\phi1}t}{\hbar}+p_1), V_{qED_{\phi_2} \rightarrow qubit}(2,t)= V_{a4}(E_{ox,2}')\sin(\frac{E_{\phi2}t}{\hbar}+p_2)$. We also notice that the time-dependent electric field coming from the EM cavity will modify the potential barrier between nodes 1 and 2 of the position-based qubit. At the first level of approximation, such modification of potential barrier can be expressed in a modification of hopping coefficients by linear renormalization given as
$t_{s12,mod, E_{\phi1}}(t)=t_s(t) exp(-c_0(V_{a5}(\frac{E_{ox1,1}'+E_{ox1,2}'}{2})\sin(\frac{E_{\phi1}t}{\hbar}+p_1))$ and $t_{s12,mod, E_{\phi2}}(t)=t_s(t) exp(-c_0(V_{a5}(\frac{E_{ox2,1}'+E_{ox2,2}'}{2})\sin(\frac{E_{\phi2}t}{\hbar}+p_2))=P(t)t_s(t)$. 
Alternative to the renormalization of hopping parameter can be done by a non-linear renormalization expressed as $t_{s12,mod}(t)=t_s(t) exp(-c_0(V_{a5}((|c_1(t)|^2+ |c_2(t)|^2)E_{o1,2}')\sin(\frac{E_{\phi1}t}{\hbar}+p_1)+(|c_3(t)|^2+ |c_4(t)|^2)V_{a6}(E_{ox,2}') \sin(\frac{E_{\phi2}t}{\hbar}+p_2)))$. However, such renormalization leads to the fact that the Hamiltonian depends on its eigenstates, and the eigenstates depend on the Hamiltonian so one needs to continue the self-consistency in calculations. Therefore, it is preferable to use a linear renormalization for hopping terms. However, both types of renormalization for hopping terms are possible. We will favor the linear renormalization of hopping parameters due to its mathematical simplicity. However, it seems that non-linear renormalization of hopping parameters
is closer to the physical realism.

Via the modifications of the tight-binding model in terms of qubit confinement Hamiltonian, we arrive at the following structure of non-interacting part of qEC-qubit Hamiltonian given as
\begin{eqnarray}
\hat{H}_{(\ket{E_{\phi}}\ket{x})_{r2,0}}=(\hat{H}_{qEC}\times \hat{I}_{qubit})_{r2,0}+\hat{I}_{EC} \times \hat{V}(t)_{eff(qED \rightarrow qubit)}+(\hat{I}_{qEC} \times \hat{H}_{qubit})_{r2,0} = 
\begin{pmatrix}
E_{\phi1} & 0 & 0 & 0 \\
0 & E_{\phi1} & 0 & 0 \\
0 & 0 & E_{\phi2} & 0 \\
0 & 0 & 0 & E_{\phi2} \\
\end{pmatrix}
+\nonumber \\
\begin{pmatrix}
V_{a1}(E_{ox,1}')\sin(\frac{E_{\phi1}t}{\hbar}+p_1) & 0 & 0 & 0 \\
0 & V_{a2}(E_{ox,1}')\sin(\frac{E_{\phi1}t}{\hbar}+p_1) & 0 & 0 \\
0 & 0 &V_{a1}(E_{ox,2}')\sin(\frac{E_{\phi2}t}{\hbar}+p_2) & 0 \\
0 & 0 & 0 & V_{a2}(E_{ox,2}')\sin(\frac{E_{\phi2}t}{\hbar}+p_2) \\
\end{pmatrix}_{r2,0}+\nonumber \\
\end{eqnarray}
\begin{eqnarray*}
\begin{pmatrix}
E_{p1}(t) & t_s(t)e^{is_0 \int_{0}^{t} dt'\bra{E_{\phi1}} \hat{E}_{f}(x_{q},t')\ket{E_{\phi1}}} & 0 & 0 \\
t_s(t)^{*}e^{-is_0 \int_{0}^{t} dt'\bra{E_{\phi1}} \hat{E}_{f}(x_{q},t')\ket{E_{\phi1}}} & E_{p2}(t) & 0 & 0 \\
0 & 0 & E_{p1}(t) & t_s(t)e^{is_0 \int_{0}^{t} dt'\bra{E_{\phi2}} \hat{E}_{f}(x_{q},t')\ket{E_{\phi2}}} \\
0 & 0 & t_s(t)^{*}e^{-is_0 \int_{0}^{t} dt'\bra{E_{\phi2}} \hat{E}_{f}(x_{q},t')\ket{E_{\phi2}}} & E_{p2}(t) \\
\end{pmatrix}_{r2,0}.
\end{eqnarray*}
\normalsize
Introducting Hamitlonians $\hat{H}_{eff1}$ and $\hat{H}_{eff2}$, we obtain $\hat{H}_{eff1}=$
\small
\begin{eqnarray*}
\begin{pmatrix}
E_{p1}(t) & e^{-c_0(V_{a5}(\frac{E_{ox1,1}'+E_{ox1,2}'}{2})\sin(\frac{E_{\phi1}t}{\hbar}+p_1)}t_s(t)e^{is_0 \int_{0}^{t} dt'\bra{E_{\phi1}} \hat{E}_{f}(x_{q},t')\ket{E_{\phi1}}} \\
 e^{-c_0(V_{a5}(\frac{E_{ox1,1}'+E_{ox1,2}'}{2}\sin(\frac{E_{\phi1}t}{\hbar}+p_1))} t_s(t)^{*}e^{-is_0 \int_{0}^{t} dt'\bra{E_{\phi1}} \hat{E}_{f}(x_{q},t')\ket{E_{\phi1}}} & E_{p2}(t) \\
\end{pmatrix}.
\end{eqnarray*}
\normalsize
and $\hat{H}_{eff2}=$
\small
\begin{eqnarray*}
\begin{pmatrix}
E_{p1}(t) & e^{-c_0(V_{a5}(\frac{E_{ox2,1}'+E_{ox2,2}'}{2})\sin(\frac{E_{\phi2}t}{\hbar}+p_2)}t_s(t)e^{is_0 \int_{0}^{t} dt'\bra{E_{\phi2}} \hat{E}_{f}(x_{q},t')\ket{E_{\phi2}}} \\
 e^{-c_0(V_{a5}(\frac{E_{ox2,1}'+E_{ox2,2}'}{2}\sin(\frac{E_{\phi2}t}{\hbar}+p_2))} t_s(t)^{*}e^{-is_0 \int_{0}^{t} dt'\bra{E_{\phi2}} \hat{E}_{f}(x_{q},t')\ket{E_{\phi2}}} & E_{p2}(t) \\
\end{pmatrix}.
\end{eqnarray*}
\normalsize
The alternative stage of renormalization of qubit Hamiltonian is due to the modifications of hopping parameters as by $t_s(t) \rightarrow t_{s12,mod}(t)$.
In such a case, we obtain alternative definitions of $\hat{H}_{eff1}$ and $\hat{H}_{eff2}$ given as
\small
\begin{eqnarray*}
\hat{H}_{eff1}=
\begin{pmatrix}
E_{p1}(t) & P(t)t_s(t)e^{is_0 \int_{0}^{t} dt'\bra{E_{\phi1}} \hat{E}_{f}(x_{q},t')\ket{E_{\phi1}}} \\
P(t) t_s(t)^{*}e^{-is_0 \int_{0}^{t} dt'\bra{E_{\phi1}} \hat{E}_{f}(x_{q},t')\ket{E_{\phi1}}} & E_{p2}(t) \\
\end{pmatrix}, \nonumber \\
\hat{H}_{eff2}=
\begin{pmatrix}
E_{p1}(t) & P(t)t_s(t)e^{is_0 \int_{0}^{t} dt'\bra{E_{\phi2}} \hat{E}_{f}(x_{q},t')\ket{E_{\phi2}}} \\
P(t) t_s(t)^{*}e^{-is_0 \int_{0}^{t} dt'\bra{E_{\phi2}} \hat{E}_{f}(x_{q},t')\ket{E_{\phi2}}} & E_{p2}(t) \\
\end{pmatrix}.
\end{eqnarray*}
Finally, we can write the renormalized non-interacting Hamiltonian of the qEC-qubit system as
\begin{eqnarray}
\hat{H}_{(\ket{E_{\phi}}\ket{x})_{r2,0}}+\hat{H}_{qEC-qubit}=
(\hat{H}_{qEC}\times \hat{I}_{qubit})_{r2,0}+(\hat{I}_{QEC} \times \hat{H}_{qubit})+\hat{I}_{EC} \times \hat{V}(t)_{eff(qED \rightarrow qubit)}+(\hat{I}_{qEC} \times \hat{H}_{qubit})_{r2,0} =\nonumber \\
=
\begin{pmatrix}
E_{\phi1} & 0 & 0 & 0 \\
0 & E_{\phi1} & 0 & 0 \\
0 & 0 & E_{\phi2} & 0 \\
0 & 0 & 0 & E_{\phi2} \\
\end{pmatrix}
+
\begin{pmatrix}
E_{p1}(t) & t_{s12}(t) & 0 & 0 \\
t_{s12}(t)^{*} & E_{p2}(t) & 0 & 0 \\
0 & 0 & E_{p1}(t) & t_{s12}(t) \\
0 & 0 & t_{s12}(t)^{*} & E_{p2}(t) \\
\end{pmatrix}+ \nonumber \\
+
\begin{pmatrix}
V_{a1}(E_{ox,1}')\sin(\frac{E_{\phi1}t}{\hbar}+p_1) & 0 & 0 & 0 \\
0 & V_{a1}(E_{ox,1}')\sin(\frac{E_{\phi1}t}{\hbar}+p_1) & 0 & 0 \\
0 & 0 &V_{a2}(E_{ox,2}')\sin(\frac{E_{\phi2}t}{\hbar}+p_2) & 0 \\
0 & 0 & 0 & V_{a2}(E_{ox,2}')\sin(\frac{E_{\phi2}t}{\hbar}+p_2) \\
\end{pmatrix}_{r2,0}+\nonumber \\
+
\begin{pmatrix}
0 & 0 & u_t(|a_t|^2|c_t|^2-a^{*}_td^{*}_tb_tc_t) & u_t(a_tb^{*}_t|c_t|^2-c_td_t^{*}|b_t|^2) \\
0 & 0 & u_t(d_t|a_t|^2c^{*}_t-|d_t|^2b_ta_t^{*})& u_t(-|b_t|^2|d_t|^2+b_t^{*}c_t^{*}a_td_t) \\
u^{*}_{t}(|a_t|^2|c_t|^2-a_td_tb^{*}_tc^{*}_t) & u^{*}_t(c_td_t^{*}|a_t|^2-|d_t|^2 a_tb^{*}_t) & 0 & 0 \\
u^{*}_t(|c_t|^2b_ta^{*}_t-d_tc^{*}_t|b_t|^2) & u_t^{*}(-|b_t|^2|d_t|^2+b_tc_ta_t^{*}d_t^{*}) & 0 & 0 \\
\end{pmatrix}+ \nonumber \\
+
\begin{pmatrix}
\hat{H}_{eff2} & 0_{2 \times 2} \\
0_{2 \times 2} & \hat{H}_{eff2} \\
\end{pmatrix} . \nonumber \\
\end{eqnarray}
It will be convenient to introduce the notation
$\hat{H}_s=
\begin{pmatrix}
u_t(|a_t|^2|c_t|^2-a^{*}_td^{*}_tb_tc_t) & u_t(a_tb^{*}_t|c_t|^2-c_td_t^{*}|b_t|^2) \\
u_t(+d_t|a_t|^2c^{*}_t-|d_t|^2b_ta_t^{*})& u_t(-|b_t|^2|d_t|^2+b_t^{*}c_t^{*}a_td_t) \\
\end{pmatrix}$. 
In most cases, $V_{a1}(E_{ox,1}') \approx V_{a2}(E_{ox,1}') $ and $V_{a1}(E_{ox,2}') \approx V_{a2}(E_{ox,2}') $ and with $V_{a1}(E_{ox,1}') \neq V_{a1}(E_{ox,2}')$.
In this approach, the qubit is interacting only with one electromagnetic mode of cavity. It is encoded in $u_t(1)=\lambda_0(1) E_{ox1}' e^{\frac{1}{\hbar i}-(E_{\phi1}-E_{\phi2})t}$. 
In a general case having $k$-th mode of electromagnetic field of qEC, we have
$u_t(k)=\lambda_0(k) E_{oxk}' e^{\frac{-1}{\hbar i}(E_{\phi_k}-E_{\phi_{k+1}})t}$. 
We can control the structure of this Hamiltonian by electrical signals applied to the gates of position-dependent qubit (Fig.\,2) which will have its effect on $E_{p1}(t)$ , $E_{p2}(t)$, $t_{s12}(t)$, coefficients $a(t)$ , $b(t)$ , $c(t)$ and $u(t)$.
Let us analyze the situation of qubit A and qubt B interacting with the EM cavity by Jaymes Cumming Hamiltonian (qcA-qEC-qcB) with the case when they have also mutual interaction by meas of electrostatic repulsion. It leads the quantum state spanned by eight basic states $\ket{E_{\phi1}}\ket{x_{1a}}\ket{x_{1b}}$, $\ket{E_{\phi1}}\ket{x_{1a}}\ket{x_{2b}}, \ket{E_{\phi1}}\ket{x_{2a}}\ket{x_{1b}}$, $\ket{E_{\phi1}}\ket{x_{2a}}\ket{x_{2b}}$, $\ket{E_{\phi2}}\ket{x_{1a}}\ket{x_{1b}}$, $\ket{E_{\phi2}}\ket{x_{1a}}\ket{x_{2b}}, \ket{E_{\phi2}}\ket{x_{2a}}\ket{x_{1b}}$, $\ket{E_{\phi2}}\ket{x_{2a}}\ket{x_{2b}}$ so we have
\begin{equation}
\ket{\psi}_t=\gamma_1(t)\ket{E_{\phi1}}\ket{x_{1a}}\ket{x_{1b}}+..+\gamma_8(t)\ket{E_{\phi2}}\ket{x_{2a}}\ket{x_{2b}}, |\gamma_1(t)|^2+..+ |\gamma_8(t)|^2=1.
\end{equation}
By reference to Fig.\,2. we introduce the Coulomb interaction terms $E_c(1,1')=\frac{q^2}{d_{1,1'}}, E_c(1,2')=\frac{q^2}{d_{1,2'}},E_c(2,1')=\frac{q^2}{d_{2,1'}},E_c(2,2')=\frac{q^2}{d_{2,2'}}$. In addition, we will assume that the electron in qubit A moving between nodes 1 and 2 will generate a vector potential that will bring the phase imprint on qubit B and, reversely, the electron in qubit B moving between nodes 1' and 2' will generate a vector potential that will affect qubit A. It is the Aharonov-Bohm effect and moving qubits will generate the mutual phase imprints
that will contribute to decoherence time $T_2$ of qubits. Those mutual phase imprints will have a tendency to be anticorrelated since energy of magnetic field of such a physical system will tend to be minimized. Therefore, for q-state $\ket{E_{\phi1(2)}}\ket{1}\ket{1'}$, we will have renormalization of hopping constant of qubit A as $t_{s12,a}(t) \rightarrow t_{s12,a}(t)e^{is_{b} \int_{0}^{t} dt'\bra{x1'} \hat{H}_{B}\ket{x2'}}$ and at the same time renormalization of the hopping constant of qubit B as $t_{s1'2',b}(t) \rightarrow t_{s1'2',a}(t)e^{is_{a} \int_{0}^{t} dt'\bra{x1} \hat{H}_{A}(t')\ket{x2}}$. It is an open issue whether we shall chose $t_{s12,a}(t) \rightarrow t_{s12,a}(t)e^{is_{b} \int_{0}^{t} dt'\bra{x1'} \hat{H}_{B}\ket{x2'}}$ or $t_{s12,a}(t) \rightarrow t_{s12,a}(t)e^{is_{b} \int_{0}^{t} dt' \sqrt{\bra{x1'} \hat{H}_{B}\ket{x2'}}}$.
In the case of qcA-qEC-qcB, in most simplistic approach we are given an 8$\times$8 system Hamiltonian that is after 3-rd renormalization step and is of the following form
\begin{eqnarray*}
\hat{H}_{(\ket{E_{\phi}}\ket{x})_{r2,0},a-b}+\hat{H}_{A-B}+\hat{H}_{qEC-qubit}=
(\hat{H}_{qEC}\times \hat{I}_{qubit,a}\times \hat{I}_{qubit,a})_{r2,0}+(\hat{I}_{QEC} \times \hat{H}_{qubit,a}\times \hat{I}_{qubit,a})+(\hat{I}_{QEC} \times \hat{I}_{qubit,a}\times \hat{H}_{qubit,a})+
 \nonumber \\+ \hat{I}_{EC} \times \hat{H}_{A-B}+\hat{I}_{EC} \times \hat{V}(t)_{eff(qED \rightarrow qubit A)} \times \hat{I}_{qubit,B} + \hat{I}_{EC} \times \hat{I}_{qubit,A} \times \hat{V}(t)_{eff(qED \rightarrow qubit B)}+ \nonumber \\ + (\hat{I}_{qEC} \times \hat{H}_{qubit A} \times \hat{I}_{qubit,B})_{r2,0}+ (\hat{I}_{qEC} \times \hat{I}_{qubit A} \times \hat{H}_{qubit,B})_{r2,0}=\nonumber \\
=
\begin{pmatrix}
E_{\phi1} & 0 & 0 & 0 & 0 & 0 & 0 & 0 \\
0 & E_{\phi1} & 0 & 0 & 0 & 0 & 0 & 0\\
0 & 0 & E_{\phi1} & 0 & 0 & 0 & 0 & 0 \\
0 & 0 & 0 & E_{\phi1} & 0 & 0 & 0 & 0\\
0 & 0 & 0 & 0 & E_{\phi2} & 0 & 0 & 0 \\
0 & 0 & 0 & 0 & 0 & E_{\phi2} & 0 & 0 \\
0 & 0 & 0 & 0 & 0 & 0 & E_{\phi2} & 0 \\
0 & 0 & 0 & 0 & 0 & 0 & 0 & E_{\phi2} \\
\end{pmatrix}
+ \nonumber \\
\begin{pmatrix}
E_{p1,a}(t) & 0 & t_{s12,a}(t)e^{is_{b} \int_{0}^{t} dt'\bra{x1'} \hat{H}_{B}(t')\ket{x2'}} & 0 \\
0 & E_{p1,a}(t) & 0 & t_{s12,a}(t)e^{is_{b} \int_{0}^{t} dt'\bra{x2'} \hat{H}_{B}(t')\ket{x1'}} \\
t_{s12,a}(t)^{*}e^{-is_{b} \int_{0}^{t} dt'\bra{x1'} \hat{H}_{B}(t')\ket{x2'}} & 0 & E_{p2,a}(t) & 0 \\
0 & t_{s12,a}(t)^{*}e^{-is_{b} \int_{0}^{t} dt'\bra{x2'} \hat{H}_{B}(t')\ket{x1'}} & 0 & E_{p2,a}(t)\\
\end{pmatrix} \nonumber \\ \times \hat{I}_{2 \times 2}+
\nonumber \\
+\hat{I}_{2 \times 2} \times \nonumber \\
\begin{pmatrix}
E_{p1',b}(t) & 0 & t_{s1'2',b}(t)e^{is_{b} \int_{0}^{t} dt'\bra{x1} \hat{H}_{A}(t')\ket{x2}} & 0\\
0 & E_{p1',b}(t) & 0 & t_{s1'2',b}(t)e^{is_{b} \int_{0}^{t} dt'\bra{x2} \hat{H}_{A}(t')\ket{x1}} \\
t_{s1'2',b}(t)^{*}e^{is_{b} \int_{0}^{t} dt'\bra{x1} \hat{H}_{A}(t')\ket{x2}} & 0 & E_{p2',b}(t) & 0\\
0 & t_{s1'2',b}(t)^{*}e^{is_{b} \int_{0}^{t} dt'\bra{x2} \hat{H}_{A}(t')\ket{x1}} & 0 & E_{p2',b}(t)\\
\end{pmatrix}+ \nonumber \\ +
\begin{pmatrix}
\frac{q^2}{d_{1,1'}} & 0 & 0 & 0 & 0 & 0 & 0 & 0 \\
0 & \frac{q^2}{d_{1,1'}} & 0 & 0 & 0 & 0 & 0 & 0\\
0 & 0 & \frac{q^2}{d_{1,2'}} & 0 & 0 & 0 & 0 & 0 \\
0 & 0 & 0 & \frac{q^2}{d_{1,2'}} & 0 & 0 & 0 & 0\\
0 & 0 & 0 & 0 & \frac{q^2}{d_{2,1'}} & 0 & 0 & 0 \\
0 & 0 & 0 & 0 & 0 & \frac{q^2}{d_{2,1'}} & 0 & 0 \\
0 & 0 & 0 & 0 & 0 & 0 & \frac{q^2}{d_{2,2'}} & 0 \\
0 & 0 & 0 & 0 & 0 & 0 & 0 & \frac{q^2}{d_{2,2'}} \\
\end{pmatrix}
+
\hat{I}_{2 \times 2} \times
\begin{pmatrix}
V_{a1,a}(E_{ox,1}')\sin(\frac{E_{\phi1}t}{\hbar}+p_1) & 0 \\
0 & V_{a2,a}(E_{ox,1}')\sin(\frac{E_{\phi1}t}{\hbar}+p_1) \\
\end{pmatrix}_{r2,0} \times \hat{I}_{2 \times 2} \nonumber \\
+\hat{I}_{4 \times 4} \times
\begin{pmatrix}
V_{a1,b}(E_{ox,1}')\sin(\frac{E_{\phi1}t}{\hbar}+p_1) & 0 \\
0 & V_{a2,b}(E_{ox,1}')\sin(\frac{E_{\phi1}t}{\hbar}+p_1) \\
\end{pmatrix}_{r2,0} +\nonumber \\
+
\begin{pmatrix}
0_{2 \times 2} & \hat{H}_{s,a} \\
\hat{H}_{s,a} & 0_{2 \times 2} \\
\end{pmatrix}\times \hat{I}_{2 \times 2}
+
\begin{pmatrix}
0_{4 \times 4} & \hat{I}_{2 \times 2} \times \hat{H}_{s,b} \\
\hat{I}_{2 \times 2} \times \hat{H}_{s,b} & 0_{4 \times 4} \\
\end{pmatrix}+
\begin{pmatrix}
\hat{H}_{eff2,a} & 0_{2 \times 2} \\
0_{2 \times 2} & \hat{H}_{eff2,a} \\
\end{pmatrix} \times \hat{I}_{2 \times 2}
+
\begin{pmatrix}
\hat{I}_{2 \times 2} \times \hat{H}_{eff2,b} & 0_{4 \times 4} \\
0_{4 \times 4} & \hat{I}_{2 \times 2} \times \hat{H}_{eff2,b} \\
\end{pmatrix},
 \nonumber \\
\end{eqnarray*}
where we have introduced the notation
\begin{eqnarray}
\hat{H}_{s,a}=
\begin{pmatrix}
u_{t,a}(+|a_{t,a}|^2|c_{t,a}|^2-a^{*}_{t,a}d^{*}_{t,a}b_tc_{t,a}) & u_{t,a}(+a_{t,a}b^{*}_{t,a}|c_{t,a}|^2-c_{t,a}d_{t,a}^{*}|b_{t,a}|^2) \\
u_{t,a}(+d_{t,a}|a_{t,a}|^2c^{*}_{t,a}-|d_{t,a}|^2b_{t,a}a_{t,a}^{*})& u_{t,a}(-|b_{t,a}|^2|d_{t,a}|^2+b_{t,a}^{*}c_{t,a}^{*}a_{t,a}d_{t,a}) \\
\end{pmatrix}, \nonumber \\
\hat{H}_{s,b}=
\begin{pmatrix}
u_{t,b}(+|a_{t,b}|^2|c_{t,b}|^2-a^{*}_{t,b}d^{*}_{t,b}b_tc_{t,b}) & u_{t,b}(+a_{t,b}b^{*}_{t,b}|c_{t,b}|^2-c_{t,b}d_{t,b}^{*}|b_{t,b}|^2) \\
u_{t,b}(+d_{t,b}|a_{t,b}|^2c^{*}_{t,b}-|d_{t,b}|^2b_{t,b}a_{t,b}^{*})& u_{t,b}(-|b_{t,b}|^2|d_{t,b}|^2+b_{t,b}^{*}c_{t,b}^{*}a_{t,b}d_{t,b}) \\
\end{pmatrix},
\end{eqnarray}
\normalsize
Here $u_{t,a(b)}$ refers to qubit A or to qubit B and the formula for u for given qubit was pointed by equation \ref{CenterQ}. 
The proposed renormalized qcA-qEC-qcB Hamiltonian has a very rich class of solutions and gives a possibility for studying entanglement dependence between three involved quantum systems what can be obtained using von-Neumman entropy that for the limited cases will have
analytic form expressed by elementary functions. It is quite straightforward to follow the reasoning for $N$ interacting position-based qubits with electromagnetic cavity with $k$ resonating modes. 

\section{Conclusion}

By imposing an occupancy of energetic state on one position-based qubit entangled to a radiation coming from a quantum coherent resonant cavity, we are enforcing the other qubit to change its state accordingly. It can be the base for the quantum communication and quantum internet. The generalization of the reasoning for $N$ qubits coupled to the resonant cavity as via a superconducting waveguide (that has a high quality factor) is quite straightforward. In most considerations, we need to go beyond the rotation phase approximation.
The concept of quantum internet was shown in this work. The more detailed picture requires taking into account various effects as decoherence processes that drive the quantum position-base qubit out of its coherence as well as decoherence processes that destroy the coherence of qEC (quantum Electromagnetic Cavity). It is quite important to underline that in order to bring the interaction of qEC with the position-based qubit, we need to place the position-based qubit either in the interior of qEC or in its proximity.
In the first case, bringing the position based qubit into the interior of qEC we are changing the resonant modes of the qEC and we are thus naturally bringing additional decoherence to the qEC. In the second case, in order to force the interaction between qEC and position-based qubit, we need to make a hole in the qEC wall. There is a non-zero electromagnetic radiation emitted outside from that hole, which brings the internal decoherence to the qEC. The larger the hole the stronger the interaction between the position-based qubit and qEC.
Therefore, the presented mathematical results shall be treated as a preliminary work on implementing a quantum communication with the position-based qubits.
In the conducted work, the simplistic approach is attempted as we are using a tight-binding model for the description of position-based qubits or a simplistic model for the matter-radiation interaction. This methodology shall be extended to take into account the Schr\"odinger description
of the position-based qubits as a more refined Quantum Electrodynamical Models and thus it is the subject of future work. The presented results open perspectives for implementing quantum Internet-of-Things devices. However, it shall be stressed that the conducted considerations are implementable when semiconductor qubits are quantum coherent and when the electromagnetic cavity maintains quantum coherence as well as when there is quantum interaction between position-based qubits and quantum electromagnetic cavity. It is achievable at very low temperatures as in range of 10\,mK. Quite obviously, we can extend the presented results to a quantum waveguide interacting with the position-based qubits, since the waveguide is a special case of the electromagnetic resonator.


\begin{thebibliography}{00}

\bibitem{Bashir19}
I. Bashir, M. Asker, C. Cetintepe, D. Leipold, A. Esmailiyan, H. Wang,
T. Siriburanon, P. Giounanlis, E. Blokhina, K. Pomorski, and R.~B.~Staszewski,
``A mixed-signal control core for a fully integrated semiconductor quantum computer system-on-chip,''
\emph{Proc. of IEEE European Solid-State Circuits Conf. (ESSCIRC)},
sec.~A2L-C4, pp.~125--128, 24~Sept.~2019.

\bibitem{Fujisawa} T. Fujisawa, T. Hayashi, HD Cheong, YH Jeong, and Y. Hirayama.
Rotation and phase-shift operations for a charge qubit in a double
quantum dot. Physica E: Low-dimensional Systems and Nanostructures,
21(2-4):10461052, 2004.

\bibitem{Petta} K. D. Petersson, J. R. Petta, H. Lu, and A. C. Gossard. Quantum
coherence in a one-electron semiconductor charge qubit. Phys. Rev. Lett.,
105:246804, 2010.

\bibitem{Dirk}
D. Leipold, Controlled Rabi Oscillations as foundation for entangled quantum aperture logic, Seminar
at UC Berkley Quantum Labs, 25th July 2018.

\bibitem{Panos} P.Giounanlis, E.Blokhina, K.Pomorski, D.R.Leipold, R.B.Staszewski, Modeling of Semiconductor Electrostatic Qubits Realized Through Coupled Quantum Dots,
10.1109/ACCESS.2019.2909489,IEEE Access, 2019

\bibitem{Pomorski_spie} Krzysztof Pomorski, Panagiotis Giounanlis, Elena Blokhina, Dirk Leipold, Pawel Peczkowski, Robert Bogdan Staszewski, From two types of electrostatic position-dependent semiconductor qubits to quantum universal gates and hybrid semiconductor-superconducting quantum computer, Proc. SPIE 11054, Superconductivity and Particle Accelerators 2018, 110540M, 2019.

\bibitem{Spalek}
Jozef Spalek, Wstep do fizyki materii skondensowanej, PWN, 2015.

\bibitem{Jaynes}
 E. T. Jaynes, and F. W. Cummings, Proc. IEEE51, 89(1963).

\bibitem{EntanglementMR}
Dimitris G. Angelakis, Stefano Mancini, Sougato Bose,
Steady state entanglement between hybrid light-matter qubits, arXiv:0711.1830, 2008.

\bibitem{Pomorski_compel} K.Pomorski, H.Akaike, A.Fujimaki, and K.Rusek. Relaxation method
in description of ram memory cell in rsfq computer, COMPEL, 38(1):395414, 2019.
\bibitem{Choi}
M.S.Choi, J.Yi, M.Y.Choi, J.Choi, and S.I.Lee. Quantum phase
transitions in josephson-junction chains. Phys. Rev. B, 57:R716R719, 1998.
\bibitem{QPT}
S.Sachdev. Quantum phase transitions. Cambridge Univ. Press, 2011.
\bibitem{Xu} H. Q. Xu. Method of calculations for electron transport in multiterminal quantum systems based on real-space lattice models. Phys. Rev. B, 66:165305.
\bibitem{Belzig}
D. Maile, S. Andergassen, and W. Belzig. Quantum phase transition
with dissipative frustration. Phys. Rev. B, 97, 2018.
\bibitem{PSSB2012}
K.Pomorski, P.Prokopow, Possible existence of field-induced Josephson junctions, Vol.249, No. 9, Physica Status Solidi B, 2012
\bibitem{SEL}
Krzysztof Pomorski, Panagiotis Giounanlis, Elena Blokhina, Dirk Leipold, Pawel Peczkowski, Robert Bogdan Staszewski,
Analytic view on Coupled Single-Electron Lines, ArXiv: 2674524 ,2019
\bibitem{Statistics}
C. Wetterich, Quantum mechanics from classical statistics,
Arxiv:0906.4919
\bibitem{Statistics1}
$http://math.ucr.edu/home/baez/quantropy.pdf$
\bibitem{Nbody}
Krzysztof Pomorski, Robert Bogdan Staszewski, Analytical Solutions for N-Electron Interacting System Confined in Graph of Coupled Electrostatic Semiconductor and Superconducting Quantum Dots in Tight-Binding Model with Focus on Quantum Information Processing, 22 October 2019, $https://arxiv.org/abs/1907.03180$.
\bibitem{Wilczek}
Frank Wilczek, Quantum Time Crystals, Phys. Rev. Lett. 109, 2012
\bibitem{Sacha}
Krzysztof Sacha, Jakub Zakrzewski,Time crystals: a review, Reports on Progress in Physics, Volume 81, Number 1, 2017
\bibitem{Birula}
I. Bialynicki-Birula, On the wavefunction of the photon,
Vol. 86, Acta Physical Polonica A, No. 1-2, Proceedings of the International Conference "Quantum Optics III", Szczyrk, Poland, 1993
\end{thebibliography}
\section*{Acknowledgment}
 The activity was suppored by the grant by Science Foundation Ireland under Grant 14/RP/I2921. 
I would like to thank to professor Jakub Rembielinski from University of Lodz for teaching me quantum mechanics as expressed in terms of projector operators. Special thanks are also given to Adam Bednorz from Univerity of Warsaw and to professor Andrew Mitchell from University College Dublin for lengthy discussions on tight-binding model. The assistance in picture preparation were done by Erik Staszewski from University College Dublin.


\vspace{12pt}

\end{document}